\documentclass{aastex701}
\usepackage{gensymb}
\usepackage{graphicx}
\usepackage{wrapfig}
\usepackage{sidecap}
\usepackage{float}
\usepackage{array}
\usepackage{booktabs}
\usepackage{soul}
\usepackage{tablefootnote}
\usepackage{sidecap}
\usepackage{orcidlink}
\usepackage{pdfrender}

\linenumbers

	\shorttitle{Carbon oxides on the Uranian moons}
	\shortauthors{Cartwright et al.}
	
\begin{document}

    \title{Tracing the source of carbon oxides on the large moons of Uranus}

    \correspondingauthor{Richard J. Cartwright}
			
            \author{\orcidlink{0000-0002-6886-6009} Richard J. Cartwright$^a$}
            \email{richard.cartwright@jhuapl.edu}
			\affiliation{Johns Hopkins University Applied Physics Laboratory, 11101 Johns Hopkins Rd, Laurel, MD 20723}

            \author{\orcidlink{0009-0007-3966-1980} Sasha Cryan}
            \email{sasha.cryan@universite-paris-saclay.fr}             
            \affiliation{Université Paris-Saclay, CNRS, Institut d’Astrophysique Spatiale, Orsay, France}    

            \author{\orcidlink{0000-0003-3001-9362} Rosario Brunetto}
            \email{rosario.brunetto@universite-paris-saclay.fr}             
            \affiliation{Université Paris-Saclay, CNRS, Institut d’Astrophysique Spatiale, Orsay, France}    
            
            \author{\orcidlink{0009-0008-0423-9525} Apolline Leclef}
            \email{apolline.leclef@universite-paris-saclay.fr}             
            \affiliation{Université Paris-Saclay, CNRS, Institut d’Astrophysique Spatiale, Orsay, France}

            \author{\orcidlink{0000-0003-2768-0694} Eric Quirico}
            \email{eric.quirico@univ-grenoble-alpes.fr}             
            \affiliation{Institut de Plan\'etologie et d'Astrophysique (IPAG), UMR 5274, CNRS, Universit\'e Grenoble Alpes, Grenoble, France} 

       	    \author{\orcidlink{0000-0002-6117-0164} Bryan J. Holler}
            \email{bholler@stsci.edu}            
			\affiliation{Space Telescope Science Institute, 3700 San Martin Drive, Baltimore, MD 21218}

            \author{\orcidlink{0000-0002-8296-6540} William M. Grundy}
            \email{Will.Grundy@nau.edu} 
            \affiliation{Department of Astronomy and Planetary Science, Northern Arizona University, 527 S Beaver Street, Flagstaff AZ, 86011}

			\author{\orcidlink{0000-0001-5888-4636} Tom A. Nordheim}
            \email{Tom.Nordheim@jhuapl.edu}             
			\affiliation{Johns Hopkins University Applied Physics Laboratory, 11101 Johns Hopkins Rd, Laurel, MD 20723}
            
            \author{\orcidlink{0000-0002-6036-1575} Ujjwal Raut}
            \email{ujjwal.raut@swri.org}             
			\affiliation{Southwest Research Institute, 6220 Culebra Road, San Antonio, TX 78238-5166}

            \author{\orcidlink{0000-0002-8592-0812} Matthew M. Hedman}
            \email{mhedman@uidaho.edu}             
            \affiliation{Department of Physics, University of Idaho, 375 Perimeter Drive, MS 0903, Moscow ID 83843}                                                                     
            
            \author{\orcidlink{0000-0002-1647-2358} Riley A. DeColibus}
            \email{riley.a.decolibus@jpl.nasa.gov}             
            \affiliation{Jet Propulsion Laboratory, California Institute of Technology, 4800 Oak Grove Drive Pasadena, CA 91109} 

			\author{\orcidlink{0000-0001-5048-6254} Chloe B. Beddingfield}
            \email{Chloe.Beddingfield@jhuapl.edu}             
			\affiliation{Johns Hopkins University Applied Physics Laboratory, 11101 Johns Hopkins Rd, Laurel, MD 20723}

            \author{\orcidlink{0000-0002-6220-2869} Marc Neveu}
            \email{marc.f.neveu@nasa.gov}             
            \affiliation{University of Maryland, 4296 Stadium Dr, College Park, MD 20742}
            \affiliation{Solar System Exploration Division, NASA Goddard Space Flight Center, 8800 Greenbelt Road, Greenbelt, MD 20771}

    		\author{\orcidlink{0000-0002-2161-4672} Christopher R. Glein}
            \email{christopher.glein@swri.org}             
			\affiliation{Southwest Research Institute, 6220 Culebra Road, San Antonio, TX 78238-5166}
                        
            \author{\orcidlink{0000-0002-2662-5776} Sara Faggi}
            \email{sara.faggi@nasa.gov}             
			\affiliation{Solar System Exploration Division, NASA Goddard Space Flight Center, 8800 Greenbelt Road, Greenbelt, MD 20771} 
            
            \author{\orcidlink{0000-0002-2662-5776} Geronimo L. Villanueva}
            \email{geronimo.l.villanueva@nasa.gov}             
			\affiliation{Solar System Exploration Division, NASA Goddard Space Flight Center, 8800 Greenbelt Road, Greenbelt, MD 20771}                   

            \author{\orcidlink{0000-0002-2770-7896} Noemi Pinilla-Alonso}
            \email{no155980@ucf.edu}             
            \affiliation{Instituto de Ciencias y Tecnologías Espaciales de Asturias (ICTEA), Oviedo, Asturias, Spain, ES}

            \author{\orcidlink{0000-0002-2430-0185} Stephanie M. Menten}
            \email{stephanie.menten@jhuapl.edu}             
            \affiliation{Johns Hopkins University Applied Physics Laboratory, 11101 Johns Hopkins Rd, Laurel, MD 20723}

			

	\begin{abstract}
         The Uranian moons Ariel, Umbriel, Titania, and Oberon are enriched in CO$_2$ mixed with CO, but the origin(s) of these carbon oxides, be they primarily native or radiolytic, remains uncertain. Using data collected by NIRSpec on the James Webb Space Telescope (JWST), we measured the spectral signature of CO$_2$ and other carbon oxides to help disentangle these hypotheses. Through comparison to laboratory data, we find that many of the detected spectral features are consistent with CO$_2$ ice, including $^{12}$CO$_2$ scattering peaks (4.15 -- 4.26 $\micron$), multi-lobe $^{13}$CO$_2$ bands (4.35 -- 4.43 $\micron$), and CO$_2$ biphonon and triphonon modes (4.80 -- 5.25 $\micron$). Our measurements show that CO$_2$ and CO are concentrated on the trailing hemispheres of the inner moons Ariel and Umbriel, potentially supporting a radiolytic production hypothesis, consistent with prior ground-based results. However, many of the identified spectral features are only observed in thick crystalline ice deposits measured in the laboratory, which may be difficult to form via radiolysis of carbon-bearing material mixed in icy regoliths. Similarly, the data exhibit weak 4.02 $\micron$ and 4.40 $\micron$ bands, hinting at the presence of carbonate minerals and $^{13}$CO$_2$ clathrates, respectively, possibly formed in the interiors of these moons. Furthermore, JWST has revealed that CO$_2$ is widespread at Uranus, present in its system of rings, ring moons, and irregular satellites, consistent with its largest moons accreting CO$_2$ and other carbon oxides from the Uranian subnebula. We conclude that exposed carbon oxides are potentially native, with their surface distributions shaped by charged particle irradiation and seasonal sublimation-condensation cycles. 
			\end{abstract}

        
		
		
	\section{Introduction} 

        The Voyager 2 (V2) spacecraft made a close pass of the Uranus system in 1986, returning valuable imaging data of its large and tidally-locked ``classical" Uranian satellites Miranda, Ariel, Umbriel, Titania, and Oberon \citep{smith1986voyager}. These data revealed terrains exhibiting a wide range of estimated heat fluxes pointing to periods of intense geologic activity in the past \citep{peterson2015elastic, beddingfield2015fault, beddingfield2022Arielheat, beddingfield2022Mirandaheat, beddingfield2023titaniaheat, bland2023Ariel}. These images also revealed possible sites of volatile release from within Ariel's interior, including hypothesized cryovolcanic features \citep{schenk1991fluid, beddingfield2021Arielcryo} and large-scale fissures that could be conduits to its interior, analogous to spreading centers on Earth \citep{beddingfield2025Arielspreading}. On Umbriel, a bright annulus in Wunda crater, along with other high albedo spots \citep{helfenstein1989evidence} could result from deposits of volatile ices and perhaps internally-derived salts \citep{sori2017wunda, cartwright2023Umbriel, scully2025UmbrielOberon, denton2026UmbrielWunda}. However, V2 was not equipped with a near-infrared mapping spectrometer and could not deduce the chemical species mantling putative cryovolcanic features and other geologic terrains on the surfaces of Uranus' large moons, nor the species incorporated into its system of rings and ring moons. 
        
        In the decades following the V2 flyby, these moons have been regularly observed using ground-based telescopes, primarily at near-infrared (NIR) wavelengths ($\sim$1 -- 2.5 $\micron$; e.g., \citealt{grundy2003discovery,  cartwright2023Umbriel}), revealing the presence of weak CO$_2$ ice features \citep[e.g.,][]{grundy2006distributions, cartwright2015UmoonCO2}, as well as hints of CO ice \citep[]{cartwright2022ArielCO2}. Analyses conducted by these and other prior studies demonstrated that the distribution of CO$_2$ exhibits clear spatial trends on the moons, with CO$_2$ detected primarily on their trailing hemispheres (subobserver longitudes 181 -- 360$\degree$W). CO$_2$ could not be conclusively detected on their leading hemispheres (subobserver longitudes 1 -- 180$\degree$W). Furthermore, the inner moons, Ariel and Umbriel, are enriched in CO$_2$ relative to the outer moons, Titania and Oberon. The spatial trends in the distribution of CO$_2$ have been cited as evidence for a radiolytic production cycle, whereby charged particles trapped in Uranus' magnetosphere preferentially interact with H$_2$O ice and carbonaceous material on the trailing hemispheres of the inner moons, forming CO$_2$ molecules within their regoliths \citep[e.g.,][]{gomis2005radiolyticCO2, raut2012radiolyticCO2}. Such a mechanism is perhaps analogous to hypothesized radiolytic production pathways for some of the CO$_2$ exposed on the surfaces of the icy Galilean satellites \citep[e.g.,][]{mccord1997NIMS,mccord1998Europasalts,hibbitts2000CallistoNIMS}. 
        
        On the Uranian moons, CO$_2$ has been identified through the presence of three narrow features centered near 1.966, 2.012, and 2.070 $\micron$, respectively consistent with the 2$\nu$$_1$+$\nu$$_3$, $\nu$$_1$+2$\nu$$_2$+$\nu$$_3$, and 4$\nu$$_2$+$\nu$$_3$ combination and overtone modes measured in `pure' CO$_2$ ice (i.e., mostly free of impurities), in which CO$_2$ molecules are presumably organized into a crystalline lattice exhibiting long-range order \citep[e.g.,][]{hansen1997spectral, quirico1997CO2SSHADE, gerakines2005NIRices}. This `CO$_2$ triplet band' has been detected primarily in reflectance spectra collected with the SpeX spectrograph \citep{rayner2003IRTFSpeX} on NASA's Infrared Telescope Facility (IRTF). More recent ground-based observations suggest that the CO$_2$ triplet bands are getting progressively weaker as  the Uranus system progresses toward northern summer solstice in 2050 (subobserver latitude 82$\degree$N), and the north poles of the moons are gradually exposed to sunlight \citep{cartwright2022ArielCO2}. This finding is consistent with volatile migration models that predict long-term cold traps are at low latitudes (30$\degree$S -- 30$\degree$N), where diurnal variations in heating dampen sublimation rates relative to the continually sunlit summer poles \citep{grundy2006distributions, sori2017wunda, steckloff2022exosphere, menten2024Ariel}. 

        Although ground-based data and results have provided useful information on the spectral properties and distribution of CO$_2$, the overall weakness of the CO$_2$ triplet band (band depths of $\sim$1 -- 15$\%$ of the continuum), as well as low-levels of residual contamination from telluric CO$_2$ gas features between 1.9 and 2.1 $\micron$, have constrained prior analyses and limited our understanding of the nature and origin of CO$_2$ and other carbon oxides in the Uranus system. Additionally, the factor of $\sim$1000 stronger CO$_2$ asymmetric stretching mode ($\nu$$_3$), centered near 4.27 $\micron$ in crystalline CO$_2$ ice \citep[e.g.,][]{hansen1997spectral, quirico1997CO2SSHADE, gerakines2020CO2}, is entirely obscured from the ground due to strong absorption by CO$_2$ and other gases in Earth's atmosphere. Observations made with the NIRSpec spectrograph (2.9 -- 5.1 $\micron$) on the James Webb Space Telescope (JWST) have brought CO$_2$ and other carbon oxides into focus on the Uranian moon Ariel, revealing prominent CO$_2$ scattering peaks between 4.15 and 4.26 $\micron$ and confirming the presence of CO ice, likely mixed with CO$_2$ \citep{cartwright2024ArielJWST}. Here, we report JWST/NIRSpec observations of Umbriel, Titania, and Oberon, alongside Ariel. We analyze the spectral signatures of CO$_2$ and other species on these moons to help decipher the identity, distribution, and origin of carbon oxides in the Uranus system.

	\section{Data and Methods}

    \subsection{Observations and Data Reduction} 
    
        JWST/NIRSpec \citep{jakobsen2022NIRSpec, boker2023NIRSpec} observed the leading and trailing hemispheres of Ariel, Umbriel, Titania, and Oberon between 09/02/2023 and 09/23/2023 with the G395M/F290LP grating (2.87 -- 5.25 $\micron$; resolving power, R $\sim$ 1000), as part of General Observer Program 1786 (observation details in Table \ref{Obs_details}). Each observation consisted of four dithers made with NIRSpec's integral field unit (IFU) and the NRSIRS2RAPID readout mode. To achieve comparable signal-to-noise for each observation, we utilized different exposure times to accommodate the differences in intrinsic brightness for each moon, ranging from the brightest moon Titania (Vmag $\sim$14) to the faintest Umbriel (Vmag $\sim$15).

    \begin{table}[h!]
    \centering
	\caption {JWST/NIRSpec G395M Observation details.} 
	\vspace{-0.cm}\hspace{-2.1cm}\begin{tabular}{cccccccc}
		\hline\hline       
		\begin{tabular}[c]{@{}l@{}}\hspace{-1 cm}Satellite \\ \end{tabular} &
		\begin{tabular}[c]{@{}l@{}}\hspace{-1 cm}Hemisphere \end{tabular} &
        \begin{tabular}[c]{@{}l@{}}\hspace{-1 cm}$^a$Subobserver \\ \hspace{-1 cm} Longitude ($\degree$) \end{tabular} &
        \begin{tabular}[c]{@{}l@{}}\hspace{-1 cm}Subobserver \\ \hspace{-1 cm} Latitude ($\degree$) \end{tabular} & 
		\begin{tabular}[c]{@{}l@{}}\hspace{-1 cm} UT Date \end{tabular} &
		\begin{tabular}[c]{@{}l@{}}\hspace{-1 cm} Start Time \end{tabular} &  
		\begin{tabular}[c]{@{}l@{}}\hspace{-1 cm}Total Exposure \\ \hspace{-1 cm}Time (s) \end{tabular} \\ 
		\hline
        \vspace{-0. cm} Ariel & Trailing & 293.4 & 64.5 & Sep. 06, 2023 & 19:35:03 & 3793.1  \\
        \vspace{-0. cm} & Leading & 64.0 & 64.5 & Sep. 07, 2023 & 17:32:06 & 3734.8  \\
        \vspace{-0. cm} Umbriel & Trailing & 272.2 & 64.1 & Sep. 08, 2023 & 15:15:18 & 8461.6  \\
        \vspace{-0. cm} & Leading & 119.8 & 64.1 & Sep. 06, 2023 & 21:10:19 & 8461.6 \\
        \vspace{-0. cm} Titania & Trailing & 275.1 & 63.9 & Sep. 23, 2023 & 20:13:15 & 2567.6  \\
        \vspace{-0. cm} & Leading & 102.5 & 64.1 & Sep. 10, 2023 & 01:03:46 & 2567.6 \\
        \vspace{-0. cm} Oberon & Trailing & 247.1 & 63.9 & Sep. 03, 2023 & 8:37:01 & 3152.2 \\
        \vspace{-0. cm} & Leading & 91.5 & 63.9 & Sep. 11, 2023 & 00:10:40 & 3152.2  \\                
		\hline 
    \label{Obs_details}
	\end{tabular}
    
    \vspace{-0.3 cm}\hspace{-3.3 cm} \textit{$^a$Listed values represent the mid-observation subobserver longitudes for each moon.}
    \end{table}
    
        The data were downloaded from the Mikulski Archive for Space Telescopes (data DOI: 10.17909/nj36-mt02) and then processed by the Science Calibration Pipeline v1.20.1 with CRDS context jwst$\_$1464.pmap to convert raw {\it uncal} files into {\it s3d} spectral cubes for all four dithers and for each of the eight observations \citep{Bushouse2023JWSTpipe}. Default pipeline parameters were used, including NSClean to remove the 1/$f$ pattern noise \citep{rauscherNSClean}. Extraction of spectra utilized a `template PSF-fitting' routine, following a technique leveraged by past studies that analyzed JWST/NIRSpec data \citep[e.g.,][]{souza2024JWSTTNOs,brunetto2025TNOsJWST, dePra2025JWSTTNOs,henault2025, sharkey2025JWSTirregulars,markwardt2025NtrojansJWST}. To conduct wavelength calibration, a wavelength grid was generated using the {\tt CRVAL3} and {\tt CRDELT3} header keywords, where each wavelength bin was matched to the corresponding spectral slice in the data cube. The centroid position of each target was first estimated by eye (good to within $<$1 pixel), and we used this centroid as an initial guess in all slices when performing PSF fitting (all four moons are point sources in NIRSpec spaxels, 0.1$''$ $\times$ 0.1$''$). 
         
\begin{figure}[h!]
            \centering
			\includegraphics[scale=0.9]{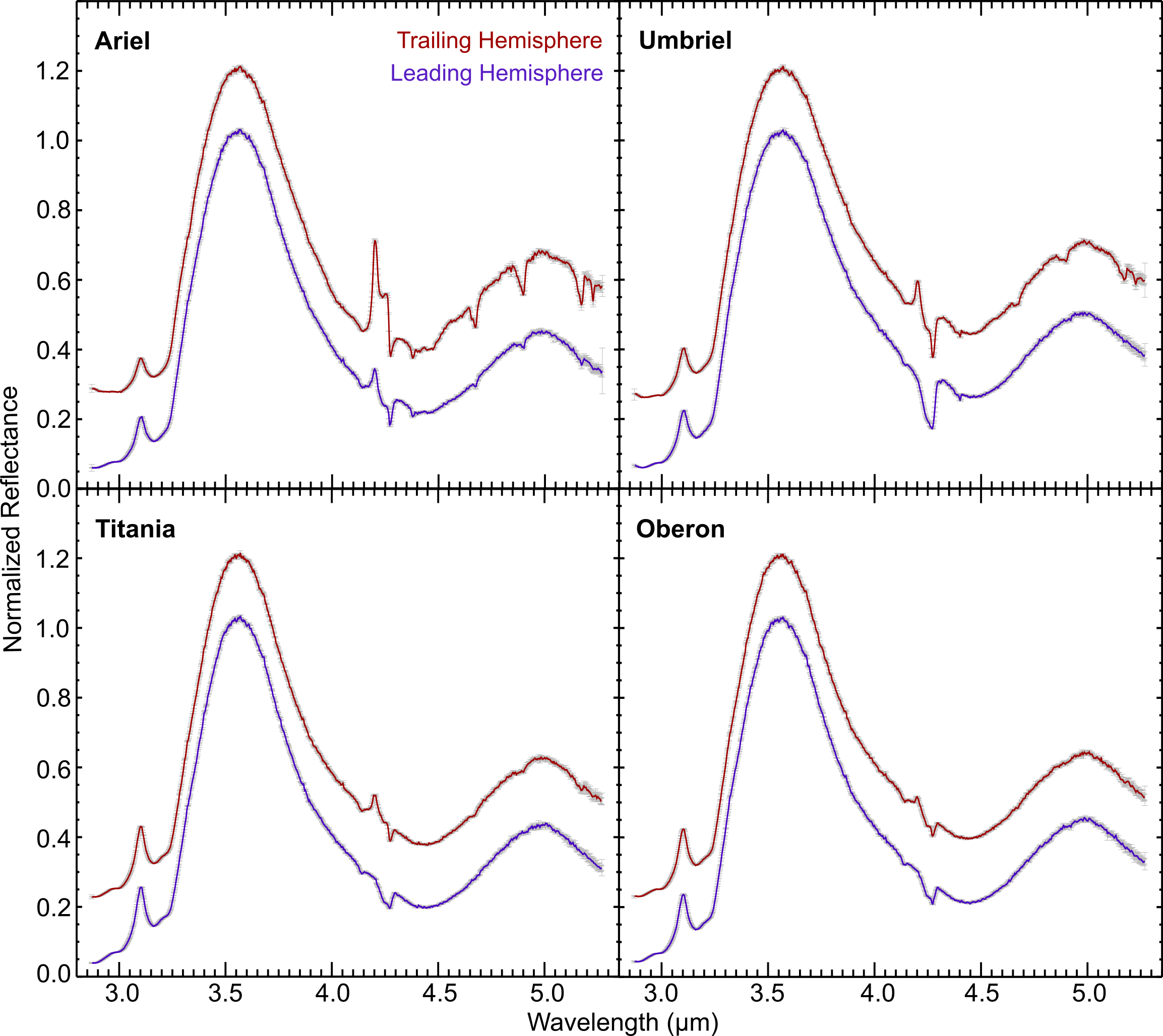}
            \centering
			\caption{\textit{NIRSpec IFU (G395M/F290LP) reflectance spectra and 1$\sigma$ uncertainties for the leading and trailing hemispheres of Ariel, Umbriel, Titania, and Oberon, normalized to one at 3.59 $\micron$ and vertically offset for clarity. The spectra of Ariel, originally reported in \citet{cartwright2024ArielJWST}, were re-reduced here using an updated version of the calibration pipeline, revealing additional spectral structure between 5.1 and 5.25 $\micron$ that was not previously presented.}}
            \centering
             \label{G395M_spectra}
		 \end{figure}

         We estimated the background by calculating the median of all spaxels more than 5 spaxels removed from the centroid, which were then subtracted from the spaxels in the slice. Using the median of a moving 21-slice window, we calculated the template PSF. Next, a 9 $\times$ 9 spaxel grid around the centroid of the template PSF was normalized to unity and then fit to the slice in the middle of the 21-slice window using the {\it scipy.optimize.minimize} function and the Nelder-Mead algorithm. We utilized the flux scaling factor and the background to make the best-fit model. The 1D target spectrum was generated by extracting the flux within a 3.5-pixel radius circular aperture, centered on each slice's centroid. We median-combined the four dithers and divided by a median G395M spectrum of P330E, a well-known calibration star \citep[G2V, Vmag 13.028 $\pm$ 0.004; e.g.,][]{bohlin2015calstars}, in order to remove the solar flux from the target spectrum. P330E spectra were constructed using the same template PSF-fitting routine described above. Uncertainties for all spectra were computed as the median absolute deviation within each wavelength bin, which were propagated to the final solar-divided target spectra (Figure \ref{G395M_spectra}). 

		 \textit{Band Measurements.} Identified spectral features were measured by first dividing each by a local linear continuum, fit to the median value of four to five data points on both sides of each feature \citep[e.g.,][]{cartwright2015UmoonCO2,cartwright2024ArielJWST}. Continuum-divided, spectral contrast measurements were made by averaging the reflectance values within $\pm$ 0.001 to 0.003 $\micron$ of the deepest/tallest point in each continuum-divided band (\textit{B$_c$}) or peak (\textit{P$_c$}), which defines the center of each spectral feature. We utilized standard error propagation procedures to estimate uncertainties for each band and peak measurement \citep[e.g.,][]{taylor1997introduction}. Next, the central wavelength position of each feature was assessed manually for quality assurance, before measuring the spectral contrasts of each absorption band (1 -- \textit{B$_d$}) and scattering peak (\textit{P$_d$} -- 1). To quantify the area of each continuum-divided feature, we used the trapezoidal rule and estimated uncertainties with Monte Carlo simulations sampling the 1$\sigma$ errors for data points within each feature. 
    
    \subsection{Laboratory Spectra} 
        
        To better understand the spectral properties of the Uranian moons and the nature of CO$_2$ on their surfaces, we compared the G395M data to laboratory experiments that measured the spectral properties of CO$_2$ ice over a range of cryogenic temperatures. We present reflectance spectra of a 10 $\pm$ 1 $\micron$ thick crystalline CO$_2$ ice layer deposited onto an organic residue and a 15 $\pm$ 1 $\micron$ thick crystalline CO$_2$ ice layer deposited directly onto the Infragold substrate, initially deposited at 50 K,  performed with the IrradiatioN de Glaces et M\'et\'eorites Analys\'ees par R\'eflectance (INGMAR), co-run by the Institut d'Astrophysique Spatiale (IAS) and the Laboratoire de physique des deux infinis Ir\`ene Joliot-Curie (IJCLab), both at the Universit\'e Paris-Saclay \citep{henault2025}. To do this, gaseous CO$_2$ was injected into a vacuum chamber (P $\sim5 \times 10^{-7}$ mbar) and deposited onto an existing $\sim$1 $\micron$ thick organic residue, produced from a previous experiment of proton-irradiated CH$_3$OH:NH$_3$ ice (subsequently warmed to room temperature), overlying an Infragold substrate initially cooled to 50 K. We also performed this experiment by condensing CO$_2$ gas directly onto the bare Infragold substrate. The CO$_2$ samples were then warmed until the CO$_2$ ice layer sublimated (130 K), with spectra collected at continuous 10 K temperature steps. Furthermore, we compared the G395M data to laboratory reflectance spectra of CO$_2$ co-deposited with H$_2$O ice ($\sim$1:1) on an Infragold substrate using INGMAR. Here, we report data collected at the initial deposition temperature of 50 K and a later 90 K temperature step.

        We also present transmission spectra of a 1 cm-thick monocrystal made of pure CO$_2$, which we compared to the G395M data to investigate weaker features between 3.3 -- 3.4 $\micron$ and 4.8 -- 5.25 $\micron$ to help determine whether these features result from thick CO$_2$ ice deposits. This crystal was grown from the gas phase under a pressure of $\sim$ 1 bar in a stainless steel cryogenic cell equipped with sapphire windows \citep[experimental setup described by][]{quirico1997near}. Spectra were measured at 187 K with a spectral resolution of 1 cm$^{-1}$. We also compared the G395M data to published laboratory spectra of CO$_2$ clathrate hydrates measured at 110 K \citep{oancea2012CO2clathrates}. Similarly, we compared the G395M data to laboratory spectra of $\sim$1 monolayer (ML) of CO$_2$ condensed on $\sim$100 nm thick meteorite simulant (amorphous MgFeSiO$_4$) and condensed on the same simulant capped by $\sim$10 to 80 ML of amorphous H$_2$O ice \citep{suhasaria2025CO2dust}. 

	\section{Results and Analyses} 
         
	\subsection{Detected Spectral Features} 

         All features detected in this study are reported in Table \ref{vibrational_modes}. Detection certainty is defined as follows: `secure' detections correspond to features observed on $\geq$2 moons at $>3$$\sigma$ level; `probable' detections correspond to features observed on 1 moon at $>3$$\sigma$ level; `tentative' detections correspond to features that have yet to be observed at $>3$$\sigma$ level and are unmeasured in this study. We also provide a compositional interpretation for all features, along with an interpretation certainty, defined as follows: `confirmed' interpretations represent features with secure detections that are very likely explained by only one compound; `favored' interpretations represent features with probable detections that are likely explained by only one compound; `ambiguous' interpretations represent all features with tentative detections. Furthermore, all features with secure or probable detections that have multiple compositional interpretations include one favored interpretation with all other interpretations listed as ambiguous.   

\begin{figure}[t!]
            \centering
			\includegraphics[scale=0.84]{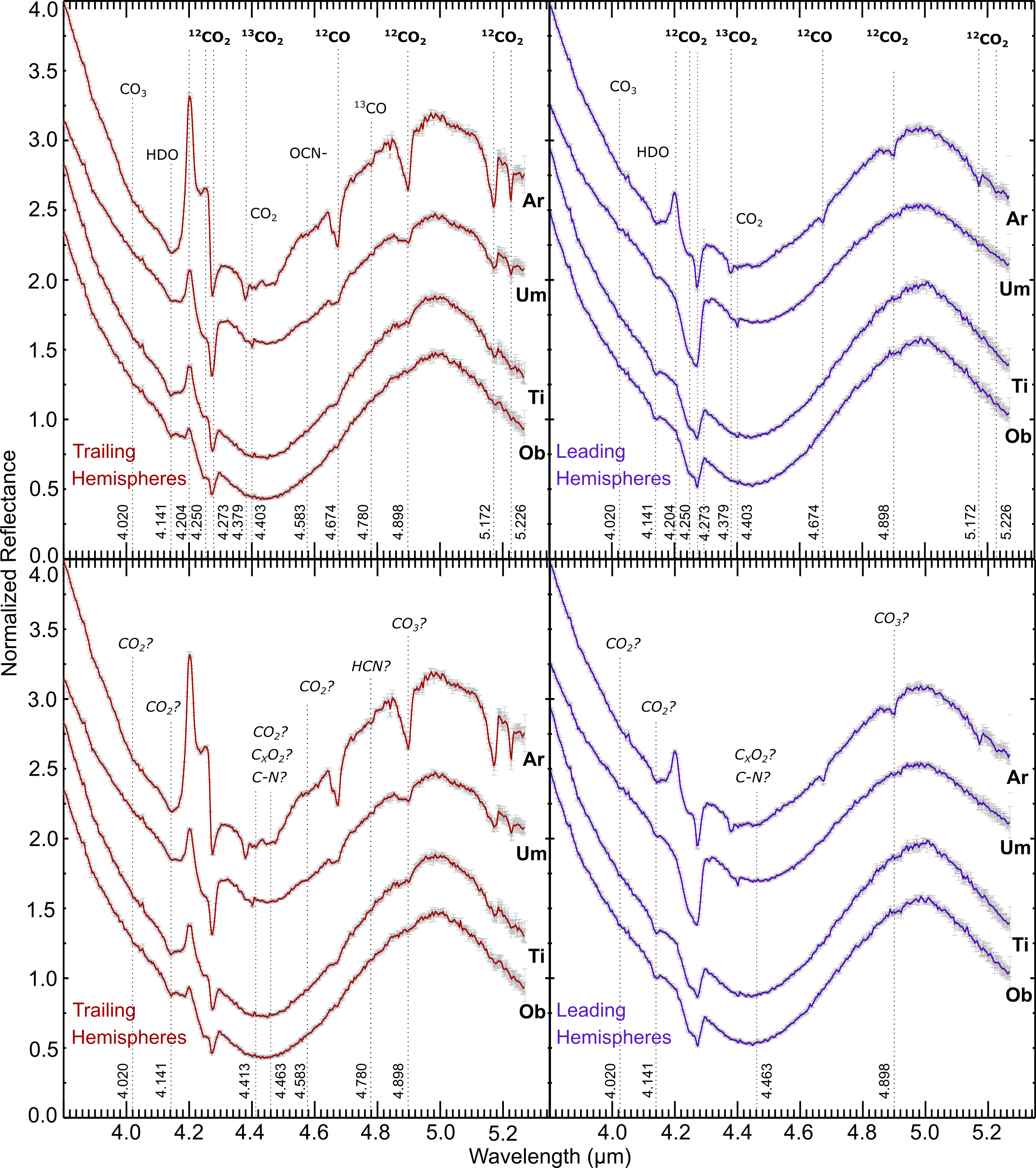}
            \centering
			\caption{\textit{NIRSpec spectra and 1$\sigma$ uncertainties for Ariel (Ar), Umbriel (Um), Titania (Ti), and Oberon (Ob), normalized to one at 4.30 $\micron$ and offset vertically for clarity (trailing hemisphere data in the left column; leading hemisphere data in the right column). The two upper panels show all features with `confirmed' (bolded text) and `favored' compositional interpretations, and the two bottom panels show all features with `ambiguous' (italicized with question marks) compositional interpretations, as described in Section 3.1 and summarized in Table \ref{vibrational_modes}. The central wavelengths ($\micron$) for all features are listed vertically along each dotted line and are included in Tables \ref{measurements_pt1}, \ref{measurements_pt2}, and \ref{measurements_pt3}.}}
            \centering
            \label{spectra_zoomin} 
		\end{figure}

\begin{SCfigure}[1.1][h!]
         \centering
			\includegraphics[width=0.7\textwidth]{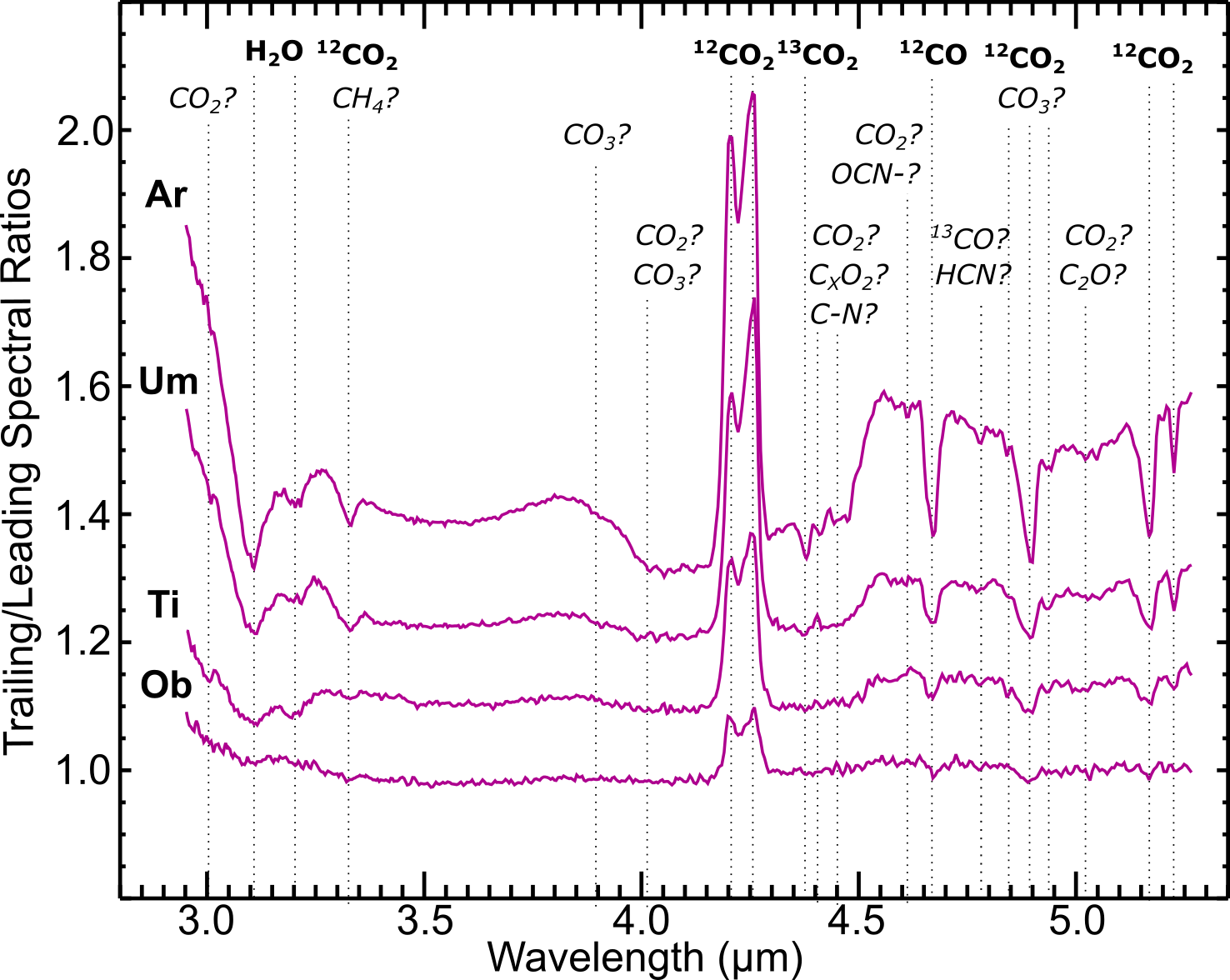}
		 \centering
			\caption{\textit{Spectral ratios generated by dividing the trailing hemisphere spectrum by the leading hemisphere spectrum for each moon, offset vertically and with uncertainties omitted for clarity. These ratios highlight the notable hemispherical asymmetries on Ariel, which become progressively weaker on Umbriel, Titania, and Oberon. The ratios also reveal subtle features near 3.01, 4.84, 4.93, and 5.02 $\micron$ that are not apparent in the non-ratio data (unmeasured, tentative detections), possibly} associated with thick CO$_2$ ice deposits ($\gtrsim$10 mm thickness; \citealt{cartwright2024ArielJWST}). An additional weak feature near 3.90 $\micron$ (unmeasured, tentative detection) in the spectral ratio of Ariel may result from carbonic acid (Table \ref{vibrational_modes}).}
        \label{spectral_ratios} 
		\end{SCfigure}

         \textit{CO$_2$ ice.} Ariel, Umbriel, Titania, and Oberon exhibit a variety of features resulting from crystalline CO$_2$ ice, in particular on their trailing hemispheres (Figures \ref{spectra_zoomin}, \ref{spectral_ratios}, and \ref{band_areas}, Table \ref{vibrational_modes}). The most pronounced examples of these CO$_2$ features are associated with the $\nu$$_3$ mode of $^{12}$CO$_2$ ($\sim$4.2 -- 4.3 $\micron$), representing one of the strongest natural absorbers measured in the laboratory \citep[e.g.,][]{dows1973CO2phonons,  Fink1982comets, gerakines2020CO2}. Scattering peaks near 4.20 $\micron$ (secure detection) are apparent in each trailing hemisphere spectrum, as well as the G395M data for Ariel's leading hemisphere, consistent with the wavelength position of a longitudinal optical (LO) phonon mode observed in crystalline $^{12}$CO$_2$ ice \citep{cooke2016CO2, suhasaria2025CO2dust}. In brief, laboratory experiments demonstrate that the $\nu$$_3$ mode of CO$_2$ ice expresses an LO mode and a transverse optical (TO) mode, whereby phonons propagate through the crystalline lattice plane parallel and plane perpendicular to incident photons, respectively. The trailing hemisphere of Ariel also exhibits a prominent peak centered near 4.25 $\micron$ (secure detection) that may represent the largest $^{12}$CO$_2$ ice Fresnel peak yet observed on an icy body (see Figure A3 in \citealt{cartwright2024ArielJWST}). Adjacent to these peaks are absorption bands centered near 4.27 $\micron$ (secure detection), likely dominated by the TO mode of $^{12}$CO$_2$ ice \cite[e.g.,][]{hansen1997spectral, quirico1997CO2SSHADE, quirico1997near, cooke2016CO2, suhasaria2025CO2dust}. The leading hemispheres of Umbriel, Titania, and Oberon do not exhibit 4.2 $\micron$ scattering peaks but do possess prominent 4.27 $\micron$ bands (Figure \ref{spectra_zoomin}). 
         
         The data display additional absorption features near 4.9 $\micron$ and 5.17 $\micron$ (both secure detections) on the trailing hemispheres of these moons and on the leading side of Ariel. These features are likely CO$_2$ biphonon plus phonon (BP+P) vibrational continua and bound triphonon (TP) modes corresponding to the ($\nu_1+\nu_2$ ; 3$\nu_2$) Fermi resonance dyad \citep{bini1991triphonons}. Briefly, Fermi resonances occur due to anharmonic interactions where the intensity of a combination or overtone mode is significantly enhanced due to interactions with a fundamental mode with similar energy and identical symmetry. 
         
         The G395M data of the inner moons, Ariel and Umbriel, exhibit additional features that are not seen on the outer moons, Titania and Oberon. Many of these features likely result from CO$_2$, including an absorption band near 4.38 $\micron$ (secure detection) that we attribute to the $\nu$$_3$ mode of $^{13}$CO$_2$ ice. Ariel's and Umbriel's 4.38 $\micron$ bands exhibit multi-lobe structures, with additional features centered near 4.40 $\micron$ (secure detection) and 4.41 $\micron$ (probable detection). We explore the spectral properties of these multi-lobe $^{13}$CO$_2$ features in section 3.3. The trailing hemispheres of Ariel and Umbriel express 5.23 $\micron$ features (secure detection) that likely represent a triphonon mode in CO$_2$ ice, resulting from collective oscillations across a thick crystalline ice lattice \citep{bini1991triphonons}. These two moons also exhibit 3.33 $\micron$ bands (secure detection) that probably stem from CO$_2$ ice, but could hypothetically include contributions from hydrocarbons, such as methane and methanol \citep[e.g.,][]{grundy2002CH4opcons, clark2009organics, dartois2010CH4clathrates, vyjidak2026CH3OHiso}. A 4.47 $\micron$ band (probable detection) detected on Ariel and possibly on Umbriel's trailing hemisphere, most likely results from thick CO$_2$ ice deposits \citep{dows1973CO2phonons}, but carbon chain oxides \citep{gerakines2001C3O2, strazzulla2007C3O2ion} or even nitrous oxide (N$_2$O) \citep[e.g.,][]{strazzulla2007C3O2ion} are possible (but less likely) alternatives. Similarly, the 4.41 $\micron$ feature could hypothetically result from nitriles (CN-bearing organics; e.g., \citealt{bernstein1997nitriles, strazzulla2007C3O2ion}), but the prevalence of other carbon oxide spectral features favors an attribution to CO$_2$.  The 4.41 and 4.47 $\micron$ bands could include contributions from $^{13}$C$^{16}$O$^{18}$O and $^{13}$C$^{18}$O$_2$, respectively \citep{bennett2010CO2CO3}. Another weak feature near 4.3 $\micron$ (probable detection) is only detected on Ariel's leading and trailing hemisphere, embedded on the long wavelength shoulder of the 4.27 $\micron$ $^{12}$CO$_2$ feature. This narrow 4.3 $\micron$ band could result from $^{12}$C$^{16}$O$^{18}$O \citep{bennett2010CO2CO3},  as well as possibly including minor contributions from amorphous CO$_2$ \citep{escribano2013amorphCO2}. Other features that may result from CO$_2$ or other carbon oxides, but are too subtle to reliably measure (identified visually, Figure \ref{spectral_ratios}), are centered near 3.01, 3.90, 4.84, 4.94, and 5.02 $\micron$ (all unmeasured, tentative detections) and discussed further in section 3.3.

         \textit{CO Ice:} The G395M data collected over the trailing hemispheres of these moons (and Ariel's leading hemisphere) show absorption bands centered near 4.67 $\micron$ (secure detection) that are consistent with the $\nu$$_3$ mode of $^{12}$CO ice \citep[e.g.,][]{gerakines2023CO}. At the peak surface temperatures of these moons (80 -- 90 K; \citealt{hanel1986infrared, sori2017wunda,menten2024Ariel}), pure CO should sublimate rapidly \citep{fray2009sublimation}, indicating it is replenished on short timescales and/or mixed with less volatile compounds. Indeed, the 4.67 $\micron$ features are flanked by weaker side bands, likely resulting from CO ice mixed with CO$_2$ ice \citep{cartwright2024ArielJWST}, as well as possibly H$_2$O ice \citep[e.g.,][]{sandford1988CO} and perhaps CO trapped as a guest molecule in a clathrate structure \citep[e.g.,][]{dartois2011CO}. A subtle feature near 4.78 $\micron$ (probable detection) on Ariel's trailing hemisphere likely results from the $\nu$$_3$ mode of $^{13}$CO ice \cite[e.g.,][]{bennett2010CO2CO3}. Although the $\nu$$_3$ mode of hydrogen cyanide (HCN) ice overlaps this wavelength range \citep{gerakines2022HCN} and could hypothetically contribute, a C-H stretching mode attributed to HCN near 3.12 $\micron$ is apparently absent, as are other HCN ice features. Because of the prevalence of other carbon oxide features in Ariel's G395M data, it seems more plausible that $^{13}$CO is the dominant contributor. Additionally, the other moons, which exhibit considerably weaker 4.67 $\micron$ $^{12}$CO features compared to Ariel, do not express 4.78 $\micron$ features, supporting our preferred interpretation of $^{13}$CO ice. For more discussion of CO ice in the Uranian system, we refer the reader to prior work that analyzed Ariel's 4.67 $\micron$ feature \citep{cartwright2024ArielJWST}. 

         \textit{CO$_3$-bearing species?} All four moons exhibit weak but broad 4.02 $\micron$ bands (secure detection) that are generally stronger on their trailing hemispheres, in particular for Ariel (Table \ref{measurements_pt1}). These broad features possibly result from a strong $\nu$$_1$ + $\nu$$_3$ combination mode exhibited by CO$_3$ in carbonate minerals \citep{hexter1958CO3, bishop2021carbnitrates}, as expressed on main belt asteroid Ceres \citep[e.g.,][]{desanctis2015ammoniated, desanctis2016bright, carrozzo2018nature} and potentially on Jupiter's moon Callisto \citep{johnson2004radiation, cartwright2024CallistoJWST}. Carbonic acid (H$_2$CO$_3$) formed via irradiation of H$_2$O and CO$_2$ mixtures, also expresses a broad feature attributed to an O-H stretching mode that could hypothetically match the 4.02 $\micron$ feature observed on the Uranian moons, but the central wavelength of this H$_2$CO$_3$ feature is usually exhibited at shorter wavelengths (3.8 -- 3.9 $\micron$), shifting in response to temperature, contaminant mixing, and other variables \citep[e.g.,][]{moore1991H2CO3, hage1998H2CO3, gerakines2000H2CO3}. Indeed, the spectral ratio of Ariel exhibits a subtle 3.9 $\micron$ feature (tentative detection) that could hypothetically result from H$_2$CO$_3$ (Figure \ref{spectral_ratios}). Irradiated CO$_2$ can also express a 4.89 $\micron$ feature attributed to the CO$_3$ molecule \citep{moll1966CO3,raut2013CO3}, which might contribute to the short wavelength wing of the Uranian moons' 4.9 $\micron$ bands \citep{cartwright2024ArielJWST}. We explore potential contributions from carbonate minerals in section 3.3. We also considered, and ruled out, possible contributions to the 4.02 $\micron$ band from sulfur-bearing species, such as sulfur dioxide (SO$_2$), because there is no reported spectral evidence at other wavelengths for S-bearing species on the Uranian moons, whose surface compositions are apparently dominated by H$_2$O ice mixed with C-bearing species \citep[e.g.,][]{grundy2006distributions}.
        
         \textit{Cyanate ion?} We note the presence of fairly weak but broad 4.59 $\micron$ shoulder features (probable detection) on the trailing hemispheres of Ariel ($>$3$\sigma$ detection) and Umbriel ($>$2$\sigma$ detection) (Figure \ref{spectra_zoomin}, Table \ref{measurements_pt2}). Similar $\sim$4.6 $\micron$ features, attributed to the cyanate ion (OCN$^-$), have been identified in JWST/NIRSpec, ground-based, and spacecraft datasets for Jupiter's moon Callisto \citep[e.g.,][]{mccord1998NIMS, cartwright2020CallistoS, cartwright2024CallistoJWST}, the interstellar medium (ISM) and protoplanetary disks \citep[e.g.,][]{pendleton1999ISMOCN-,hudson2001OCN-, mcclure2023JWSTISM, sturm2023disksJWST, potapov2025diskJWST}, young stellar objects \citep{boogert2022YSOs}, and some TNOs \citep{cryan2025TNOsnitrogen}. Laboratory experiments have shown that OCN$^-$ forms via irradiation of mixed NH$_3$, CO (and/or CO$_2$), and H$_2$O substrates at temperatures relevant to the outer Solar System and the ISM \citep[e.g.,][]{van2004OCN-,bennett2010OCN, chen2011OCN-}, and can also form thermally via interactions between H$_2$O and isocyanic acid (HNCO) at temperatures above 110 K \citep{raunier2003OCNthermal}. Spectral hints of NH$_3$, and/or NH$_4$-bearing species, have been identified previously on the Uranian moons \citep{bauer2002near,cartwright2018Umoonred,cartwright2020ArielNH3, cartwright2023Umbriel, decolibus2023MirandaNH3}, supporting the possibility that weak 4.59 $\micron$ features may result from OCN$^-$, formed via irradiation of NH-bearing species mixed with CO$_2$ and H$_2$O in their regoliths. Of note, we found no trace of a $\nu$$_3$ stretching mode expressed by NH$_3$ and other NH-bearing compounds near 2.96 $\micron$ in the G395M data, consistent with other studies that looked for this spectral feature on Ariel \citep{cartwright2024ArielJWST} and Pluto's moon Charon \citep{protopapa2024CharonJWST}. Alternatively,  perhaps the 4.59 $\micron$ band results from a poorly characterized CO$_2$ `shoulder' feature (see Fig. 2 in \citealt{cartwright2024ArielJWST}), which appears to be present in some laboratory datasets \citep{hansen1997spectral}.  A more complete exploration of OCN$^-$ vs. CO$_2$ contributions near 4.59 $\micron$ is beyond the scope of this study and left for future work.
        
         \textit{H$_2$O and HDO ice.} The Uranian moons exhibit multiple strong features indicative of crystalline H$_2$O ice. All four moons display prominent H$_2$O ice Fresnel features in the 2.90 to 3.25 $\micron$ region \citep{mastrapa2009H2Oopcon}, typified by scattering peaks near 3.10 $\micron$, attributed to the $\nu$$_3$ TO mode of H$_2$O ice, flanked by a weaker peak near 2.95 $\micron$, attributed to the $\nu$$_3$ LO mode and the $\nu$$_1$ out-of-phase mode, and another peak near 3.2 $\micron$, attributed to the $\nu$$_1$ in-phase mode \citep{bertie1964H2Omodes, whalley1977H2Omodes, mastrapa2009H2Oopcon}. The H$_2$O ice Fresnel region is stronger on the leading hemispheres of these satellites, with notably stronger 3.1 $\micron$ Fresnel peaks on the leading sides of the inner moons, Ariel and Umbriel, but only minor differences between the leading and trailing hemispheres of the outer moons, Titania and Oberon (Table \ref{measurements_pt3}). The 3.6 $\micron$ continuum peak exhibits similar hemispherical trends on these moons, with more prominent differences for the inner moons, Ariel and Umbriel, compared to the outer moons, Titania and Oberon (Table \ref{measurements_pt3}). Furthermore, the strongest H$_2$O ice features on these four moons are associated with the leading hemisphere of Titania, whereas the weakest H$_2$O ice features are associated with the trailing hemispheres of Ariel and Umbriel. Ground-based observations of the Uranian moons, made with the SpeX spectrograph on NASA’s Infrared Telescope Facility \citep{rayner2003IRTFSpeX}, also show stronger H$_2$O ice features on the leading hemispheres of these moons, but with Ariel exhibiting the strongest overall H$_2$O ice features, not Titania \citep{grundy2006distributions, cartwright2015UmoonCO2, cartwright2018Umoonred}. We discuss this disparity in section 4.2. 
        
         All of the G395M data display absorption features near 4.14 $\micron$. Similar 4.14 $\micron$ bands in Saturn's icy rings and ring moons \citep{clark2019VIMSratios, hedman2024SatringsJWST}, on icy moons at Saturn \citep{clark2019VIMSratios, brown2025SatmoonsHDO}, and putatively on Ariel \citep{cartwright2024ArielJWST} have been attributed to `semi-heavy' water in ordinary H$_2$O ice, whereby deuterium (D) substitutes for one of the hydrogen atoms in water molecules. However, the proximity of strong $^1$$^2$CO$_2$ ice scattering peaks near 4.2 $\micron$ makes assessment of putative HDO features difficult, especially on CO$_2$-rich Ariel and Umbriel where measurements of the 4.14 $\micron$ band are likely unreliable. On Titania and Oberon, however, where CO$_2$ ice features are weaker, the 4.14 $\micron$ band may be adequately isolated from nearby CO$_2$ features to permit robust measurements. On these two outer moons, we find that the 4.14 $\micron$ feature is slightly stronger on their leading sides ($<$2$\sigma$ difference). Furthermore, when comparing the moons, Titania's leading hemisphere exhibits the strongest 4.14 $\micron$ band, seemingly consistent with the stronger H$_2$O ice Fresnel region and 3.6 $\micron$ continuum peak on the leading side of this moon.  We refer the reader to prior work for deeper investigation of H$_2$O ice on the Uranian moons \citep[e.g.,][]{grundy2006distributions, cartwright2015UmoonCO2, cartwright2018Umoonred, cartwright2020UmoonIRAC}, and to a complementary study that calculates D/(D+H) ratios using the G395M data of Titania reported here, modeled with new H$_2$O ice and HDO ice optical constants measured in the laboratory \citep{tegler2026HDO}.

	\subsection{Spectral Feature Measurements} 
	     \textit{Secure detections.} The trailing hemispheres of Ariel, Umbriel, Titania, and Oberon exhibit prominent 4.02, 4.14, 4.20, 4.27, 4.67, and 4.90 $\micron$ absorption bands and 3.10 $\micron$ Fresnel peaks and 3.60 $\micron$ continuum peaks, with $>$3$\sigma$ detection of both band depths or peak heights (`spectral contrast') and areas for these features (Tables \ref{measurements_pt1} and \ref{measurements_pt3}). The trailing hemispheres of Ariel, Umbriel, and Titania also exhibit prominent 5.17 $\micron$ features ($>$3$\sigma$ detection), whereas this feature is slightly weaker on Oberon's trailing side ($>$2$\sigma$). Similarly, the leading hemisphere of Ariel exhibits all six of these features ($>$3$\sigma$), but they are considerably weaker than the same features on its trailing hemisphere. The leading hemispheres of the other three moons exhibit strong 4.27 $\micron$ bands ($>$3$\sigma$), and Titania ($>$3$\sigma$), Umbriel ($>$2$\sigma$), and Oberon ($>$2$\sigma$) exhibit broad 4.02 $\micron$ bands as well. The leading sides of Umbriel and Titania also express weak 4.67 $\micron$ features ($>$2$\sigma$). The trailing hemispheres of Ariel and Umbriel express absorption features centered near 3.33, 4.38, 4.40, and 5.23 $\micron$ ($>$3$\sigma$). The leading hemisphere of Ariel exhibits 4.38 $\micron$ and 4.4 $\micron$ features ($>$3$\sigma$). The 4.20 and 4.25 $\micron$ features are convolved, and we report the same area for both features. 
         
         \textit{Probable detections.} Weaker features near 4.30, 4.41, 4.47, 4.59, and 4.78 $\micron$ are detected on the trailing hemisphere of Ariel, and the 4.30, 4.41, 4.47 $\micron$ features ($>$3$\sigma$) and a 4.59 $\micron$ band ($>$2$\sigma$) are also present on Ariel's leading side (Table \ref{measurements_pt2}). Weak 4.41, 4.47, and 4.59 $\micron$ features are present on Umbriel's trailing hemisphere but are detected at a lower confidence level ($>$2$\sigma$), limiting robust measurements of these three features to only one moon (Ariel), making their detection `probable' and not `secure' (see opening paragraph of Section 3.1). The 4.40 and 4.41 $\micron$ features are convolved, and we report the same area for both.

\begin{SCfigure}[1.1]
            \centering
			\includegraphics[width=0.72\textwidth]{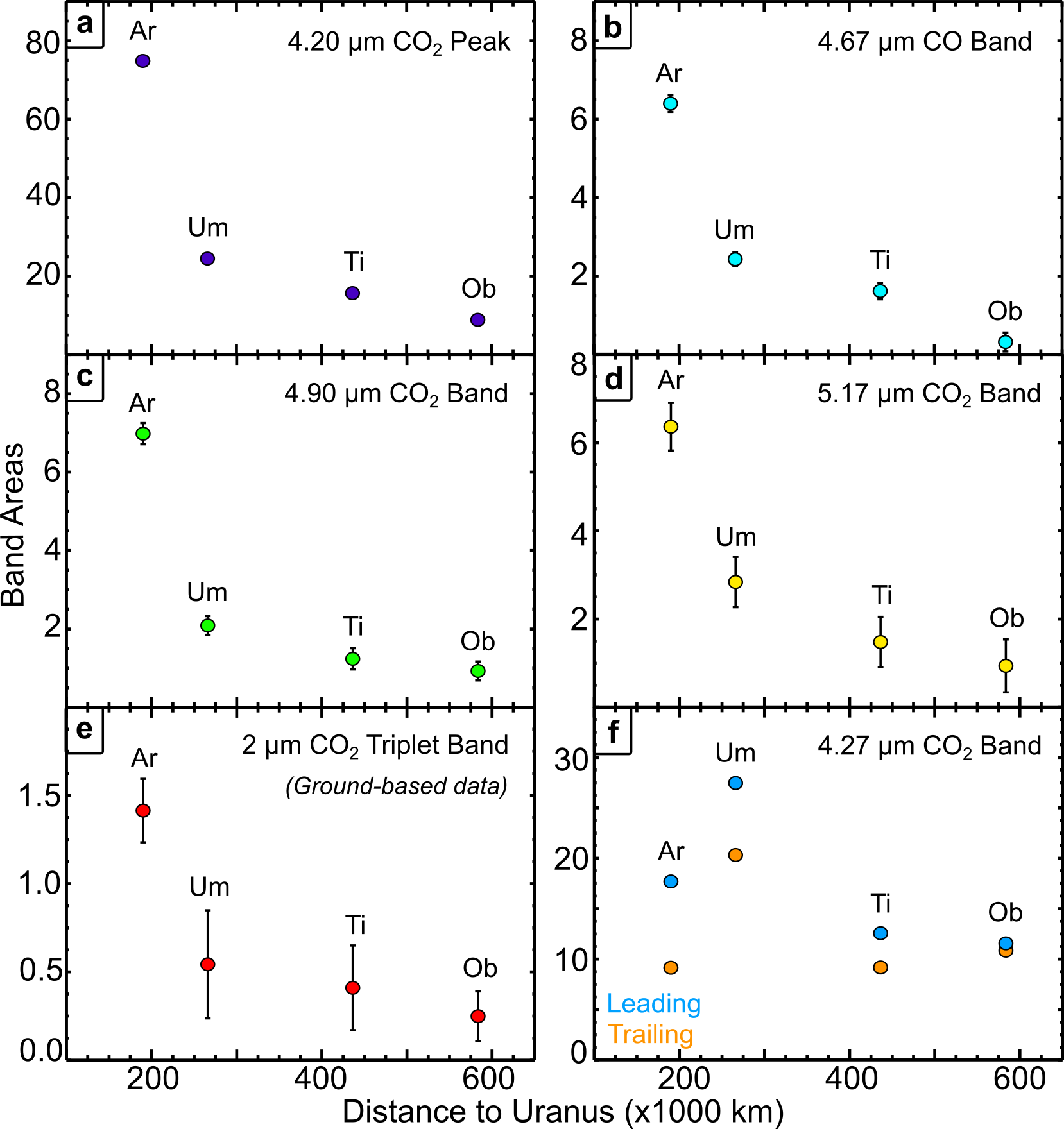}          
			\caption{\textit{Area measurements and 3$\sigma$ uncertainties for the (a) 4.20 $\micron$ CO$_2$ peak, (b) 4.67 $\micron$ CO band, (c) 4.90 $\micron$ CO$_2$ band, (d) 5.17 $\micron$ CO$_2$ band, (e) the `2 $\micron$' CO$_2$ triplet band \citep[e.g.,][]{grundy2006distributions}, and (f) the 4.27 $\micron$ CO$_2$ band (in some cases, the 3$\sigma$ uncertainties are smaller than the symbols shown here). All measurements were scaled up by a factor of 1000 (Table \ref{measurements_pt1}). For (a-e), only trailing hemisphere measurements are shown as most non-Ariel leading hemisphere measurements are ambiguous ($<$3$\sigma$ detection). For the strong 4.27 $\micron$ band (f), measurements of both the leading and trailing hemispheres were made for all four moons. The notable disparity in the apparent trends displayed by the 4.27 $\micron$ band measurements compared to the other spectral features highlights the challenges and uncertainties in interpreting this feature (see Section 4.1).}}
            \centering
            \label{band_areas}
		\end{SCfigure}

         Thus, Ariel's trailing hemisphere exhibits a rich variety of prominent and subtle spectral features,  while its leading hemisphere, and the trailing hemisphere of Umbriel, express weaker versions of many of the same features. The trailing hemispheres of Titania and Oberon exhibit a small number of these spectral features, whereas their leading sides only exhibit 4.02 $\micron$ and 4.27 $\micron$ bands. We compare band and peak area measurements for the most prominent CO$_2$ (4.20 $\micron$, 4.27 $\micron$, 4.90 $\micron$, and 5.17 $\micron$) and CO (4.67 $\micron$) features to ground-based measurements of the CO$_2$ triplet band to discern relevant trends in the distribution of these species (Figure \ref{band_areas}) and discuss the measurements further in section 4.1.

        \begin{table}[h!]
        \centering
    	\caption {Confirmed and suspected carbon oxide spectral features.} 
    	\vspace{-0.cm}\hspace{-1.3 cm}\begin{tabular}{ccccc}
    		\hline\hline       
    		\begin{tabular}[c]{@{}l@{}}\hspace{-1 cm}Spectral Feature \end{tabular} &
    		\begin{tabular}[c]{@{}l@{}}\hspace{-1 cm}Detection \\ \hspace{-1 cm}Certainty \end{tabular} &        
            \begin{tabular}[c]{@{}l@{}}\hspace{-1 cm}Interpretation(s) \end{tabular} & 
            \begin{tabular}[c]{@{}l@{}}\hspace{-1 cm}Interpretation \\ \hspace{-1 cm}Certainty \end{tabular} &             
            \begin{tabular}[c]{@{}l@{}}\hspace{-1 cm}Vibrational \\ \hspace{-1 cm}Modes \end{tabular} \\ 
    		\hline
            \vspace{-0. cm}  3.01 $\micron$ band & Tentative  & \textit{CO$_2$?} & Ambiguous & Unknown$^a$ \\
            \textit{(not measured)} & & & & \\
            \hline
            \vspace{-0. cm} 3.33 $\micron$ band & Secure & $^{12}$CO$_2$ & Favored & Unknown$^a$ (\textit{$\nu$$_2$+$\nu$$_3$ mag. dipole?}; \\ 
             & & & & \citealt{kazakov2021CO2gasM1}) \\
            & & \textit{Hydrocarbons?}& Ambiguous & $\nu$$_3$ \citep[e.g.,][]{grundy2002CH4opcons} \\
            \hline 
            \vspace{-0. cm} 3.90 $\micron$ band & Tentative & \textit{H$_2$CO$_3$?} & Ambiguous & $\nu$(OH) \citep[e.g.,][]{moore1991H2CO3} \\
            \textit{(not measured)} & & & & \\
            \hline            
            \vspace{-0. cm} 4.02 $\micron$ band & Secure & CO$_3$ & Favored & $\nu$$_1$+$\nu$$_4$ \citep{hexter1958CO3} \\
            & & \textit{H$_2$CO$_3$?} & Ambiguous & $\nu$(OH) \citep[e.g.,][]{moore1991H2CO3} \\
            & & \textit{CO$_2$?} & Ambiguous & Christiansen feature \citep{dePra2025JWSTTNOs} \\
            \hline
            \vspace{-0. cm} 4.20 $\micron$ peak & Secure & \textpdfrender{TextRenderingMode=FillStroke,LineWidth=0.5pt}{$^{12}$CO$_2$} & Confirmed & $\nu$$_3$$^a$ \\
            \hline
            \vspace{-0. cm} 4.25 $\micron$ peak & Secure & \textpdfrender{TextRenderingMode=FillStroke,LineWidth=0.5pt}{$^{12}$CO$_2$} & Confirmed & $\nu$$_3$$^a$ \\
            \hline
            \vspace{-0. cm} 4.27 $\micron$ band & Secure & \textpdfrender{TextRenderingMode=FillStroke,LineWidth=0.5pt}{$^{12}$CO$_2$} & Confirmed & $\nu$$_3$$^a$ \\
            \hline
            \vspace{-0. cm} 4.30 $\micron$ band & Probable & $^{12}$C$^{16}$O$^{18}$O & Favored & $\nu$$_3$ \citep{bennett2010CO2CO3} \\ 
            & & \textit{Amorphous $^{12}$CO$_2$?} & Ambiguous  & $\nu$$_3$ (\citealt{escribano2013amorphCO2}) \\
            \hline
            \vspace{-0. cm} 4.38 $\micron$ band & Secure & \textpdfrender{TextRenderingMode=FillStroke,LineWidth=0.5pt}{$^{13}$CO$_2$} & Confirmed &$\nu$$_3$$^a$ \\
            \hline
            \vspace{-0. cm} 4.40 $\micron$ band & Secure & $^{13}$CO$_2$ clathrates & Favored &  $\nu$$_3$ \citep{oancea2012CO2clathrates} \\
            & & \textit{$^{13}$CO$_2$?} & Ambiguous & $\nu$$_3$ (This work, Figure \ref{CO2_4.0-4.6}) \\
            \hline
            \vspace{-0. cm} 4.41 $\micron$ band & Probable & \textit{$^{13}$CO$_2$?} & Ambiguous & $\nu$$_3$ (This work, Figure \ref{CO2_4.0-4.6}) \\
            & & \textit{$^{13}$C$^{16}$O$^{18}$O?} & Ambiguous & $\nu$$_3$ \citep{loeffler2005CO}) \\ 
            & & \textit{C$_X$O$_2$?} & Ambiguous & $\nu$$_3$ \citep{gerakines2001C3O2} \\
            & & \textit{Nitriles?} & Ambiguous & $\nu$$_3$ \citep[e.g.,][]{strazzulla2007C3O2ion} \\
            \hline
            \vspace{-0. cm} 4.47 $\micron$ band & Probable & \textit{$^{13}$C$^{18}$O$_2$?} & Ambiguous & $\nu$$_3$ \citep{loeffler2005CO} \\
            & & \textit{Nitriles?} & Ambiguous & $\nu$$_3$ \citep[e.g.,][]{strazzulla2007C3O2ion} \\
            & &  \textit{C$_X$O$_2$?} & Ambiguous & $\nu$$_3$ \citep{gerakines2001C3O2} \\
            \hline
            \vspace{-0. cm} 4.59 $\micron$ band & Probable & OCN$^-$ & Ambiguous & $\nu$$_3$ \citep{grim1987OCN-}\\
            & &  \textit{C$_X$O$_2$?} & Ambiguous & $\nu$$_3$ \citep{gerakines2001C3O2} \\
            & & \textit{CO$_2$?} & Ambiguous & Unknown$^a$ \\
            \hline
            \vspace{-0. cm} 4.67 $\micron$ band & Secure & \textpdfrender{TextRenderingMode=FillStroke,LineWidth=0.5pt}{$^{12}$CO} & Confirmed & $\nu$$_3$ \citep[e.g.,][]{ewing1961COspectro} \\ 
            \hline            
            \vspace{-0. cm} 4.78 $\micron$ band & Probable & $^{13}$CO & Favored & $\nu$$_3$ \citep[e.g.,][]{ewing1961COspectro} \\ 
            & &  \textit{HCN?} & Ambiguous & $\nu$$_3$ \citep{gerakines2022HCN} \\        
            \hline            
            \vspace{-0. cm} 4.84 $\micron$ band & Tentative & \textit{CO$_2$?} & Ambiguous & Triphonon$^{a,b}$ \\ 
            \textit{(not measured)} & & & & \\
            \hline
            \vspace{-0. cm} 4.90 $\micron$ band & Secure & \textpdfrender{TextRenderingMode=FillStroke,LineWidth=0.5pt}{CO$_2$} & Confirmed & Biphonon + phonon$^{a,b}$ \\
            & & \textit{CO$_3$?} & Ambiguous & \citep{raut2013CO3} \\
            \hline
            \vspace{-0. cm} 4.94 $\micron$ band & Tentative & \textit{$^{12}$C$^{16}$O$^{18}$O?} & Ambiguous &  $^{a,b}$ \\ 
            \textit{(not measured)} & & & &  \\
            \hline
            \vspace{-0. cm} 5.02 $\micron$ band & Tentative & \textit{CO$_2$?} & Ambiguous & Unknown$^a$ \\
            \textit{(not measured)} & & \textit{C$_2$O?} & Ambiguous & \citep[e.g.,][]{bennett2010CO2CO3} \\
            \hline
            \vspace{-0. cm} 5.17 $\micron$ band & Secure & \textpdfrender{TextRenderingMode=FillStroke,LineWidth=0.5pt}{CO$_2$} & Confirmed & Biphonon + phonon$^{a,b}$ \\ 
            \hline
            \vspace{-0. cm} 5.23 $\micron$ band & Secure & \textpdfrender{TextRenderingMode=FillStroke,LineWidth=0.5pt}{CO$_2$} & Confirmed & Triphonon$^{a,b}$ \\  
    		\hline 
        \label{vibrational_modes}
    	\end{tabular}

        \vspace{-0.3 cm}\hspace{-1.6 cm} \textit{$^a$Interpreted CO$_2$ features identified using data reported in \citealt{hansen1997spectral,quirico1997CO2SSHADE, quirico1997near}.}
        
        \hspace{0.1 cm}$^b$Features between 4.8 and 5.25 $\micron$ likely result from the \textit{$\nu$$_2$+$\nu$$_3$ ; 3$\nu$$_2$} Fermi resonance region \citep{bini1991triphonons}.
        \end{table}
    
    \begin{table}[t]
    \centering
	\caption {Measurements of carbon oxide spectral features on Ariel, Umbriel, Titania, and Oberon.} 
	\vspace{-0.4cm}\begin{tabular}{cccccccc}
		\hline\hline
		\begin{tabular}[c]{@{}l@{}}\hspace{-1 cm}Feature \\ \hspace{-1 cm}Name \end{tabular} & 
		\begin{tabular}[c]{@{}l@{}}\hspace{-1 cm}Satellite \\ \end{tabular} &
		\begin{tabular}[c]{@{}l@{}}\hspace{-1 cm}Hemisphere \end{tabular} &         
		\begin{tabular}[c]{@{}l@{}}\hspace{-1 cm}Feature \\ \hspace{-1 cm}Center ($\micron$) \end{tabular} &  
		\begin{tabular}[c]{@{}l@{}}\hspace{-1 cm}Wavelength \\ \hspace{-1 cm} Range ($\micron$) \end{tabular} & 
		\begin{tabular}[c]{@{}l@{}}\hspace{-1 cm}Spectral \\ \hspace{-1 cm}Contrast ($\%$) \end{tabular} &
		\begin{tabular}[c]{@{}l@{}}\hspace{-1 cm}Band Area \\ \hspace{-1 cm}(10$^-$$^3$ $\micron$) \end{tabular} &		
        \begin{tabular}[c]{@{}l@{}}\hspace{-1 cm}$>$3$\sigma$ Spec. Contrast \\ \hspace{-1 cm}\& Band Area? \end{tabular} \\
		\hline

        \vspace{-0. cm}4.02 $\micron$ & Ariel & Trailing & 4.02 & 3.96 -- 4.11 & 7.83 $\pm$ 0.58 & 59.18 $\pm$ 0.82 & Yes \\
        Band & & Leading & '' & '' & 3.97 $\pm$ 0.33 & 30.12 $\pm$ 0.77 & Yes \\
		\vspace{-0. cm}& Umbriel & Trailing & '' & '' & 1.82 $\pm$ 0.47 & 7.75 $\pm$ 0.61 & Yes \\
          & & Leading & '' & '' & 1.70 $\pm$ 0.56 & 0.28 $\pm$ 0.67 & No \\
		\vspace{-0. cm}& Titania & Trailing & '' & '' & 2.42 $\pm$ 0.42 & 16.75 $\pm$ 0.64 & Yes \\
          & & Leading & '' & '' & 2.44 $\pm$ 0.52 & 11.47 $\pm$ 0.72 & Yes \\
		\vspace{-0. cm}& Oberon & Trailing & '' & '' & 2.16 $\pm$ 0.53 & 9.35 $\pm$ 0.81 & Yes \\
          & & Leading & '' & '' & 1.74 $\pm$ 0.65 & 5.69 $\pm$ 0.70 & No \\  
        \hline
        
        \vspace{-0. cm}4.20 $\micron$ & Ariel & Trailing$^a$ & 4.204 & 4.152 -- 4.276 & 127.95 $\pm$ 1.11 & 74.86 $\pm$ 0.14 & Yes \\
        \vspace{-0. cm}Peak & & Trailing$^a$ & 4.252 & 4.152 -- 4.276 & 82.42 $\pm$ 1.10 & 74.86 $\pm$ 0.14 & Yes \\  
          & & Leading & '' & '' & 39.43 $\pm$ 0.82  & 19.64 $\pm$ 0.10 & Yes \\
        \vspace{-0. cm}& Umbriel & Trailing & 4.205 & 4.152 -- 4.276 & 46.88 $\pm$ 1.08 & 24.47 $\pm$ 0.08 & Yes \\
          & & Leading & - & - & - & - & No \\
        \vspace{-0. cm}& Titania & Trailing & 4.204 & 4.152 -- 4.276 & 29.29 $\pm$ 0.58 & 15.66 $\pm$ 0.05 & Yes \\
          & & Leading & - & - & - & - & No \\
        \vspace{-0. cm}& Oberon & Trailing & 4.204 & 4.152 -- 4.276 & 18.43 $\pm$ 0.82 & 8.86 $\pm$ 0.09 & Yes \\
          & & Leading & - & - & - & - & No \\          
        \hline        
        
		\vspace{-0. cm}4.27 $\micron$ & Ariel & Trailing & 4.274 & 4.258 -- 4.303 & 42.40 $\pm$ 0.51 & 9.14 $\pm$ 0.05 & Yes \\
          Band & & Leading & 4.272 & 4.200 -- 4.302 & 37.91 $\pm$ 0.85 & 17.70 $\pm$ 0.06 & Yes \\
		\vspace{-0. cm}& Umbriel & Trailing & 4.273 & 4.200 -- 4.302 & 44.58 $\pm$ 0.31 & 20.32 $\pm$ 0.06 & Yes \\
          & & Leading & 4.269 & 4.188 -- 4.302 & 48.05 $\pm$ 0.31 & 27.46 $\pm$ 0.06 & Yes \\
		\vspace{-0. cm}& Titania & Trailing & 4.273 & 4.200 -- 4.302 & 23.20 $\pm$ 0.58 & 9.17 $\pm$ 0.07 & Yes \\
          & & Leading & '' & '' & 23.66 $\pm$ 0.61 & 12.57 $\pm$ 0.06 & Yes \\
		\vspace{-0. cm}& Oberon & Trailing & '' & '' & 21.58 $\pm$ 0.63 & 10.84 $\pm$ 0.07 & Yes \\
          & & Leading & '' & '' & 23.54 $\pm$ 0.48 & 11.56 $\pm$ 0.07 & Yes \\    
        \hline
           
		\vspace{-0. cm}4.67 $\micron$ & Ariel & Trailing & 4.674 & 4.640 -- 4.706 & 23.47 $\pm$ 0.42 & 6.40 $\pm$ 0.07 & Yes \\
          Band & & Leading & '' & '' & 7.41 $\pm$ 1.14 & 1.01 $\pm$ 0.08 & Yes \\
		\vspace{-0. cm}& Umbriel & Trailing & 4.675 & '' & 7.48 $\pm$ 0.63 & 2.43 $\pm$ 0.06 & Yes \\
          & & Leading & '' & '' & 1.60 $\pm$ 0.67 & 0.38 $\pm$ 0.07 & No \\
		\vspace{-0. cm}& Titania & Trailing & 4.671 & '' & 5.26 $\pm$ 0.51 & 1.62 $\pm$ 0.07 & Yes \\
          & & Leading & '' & '' & 1.45 $\pm$ 0.56 & 0.55 $\pm$ 0.08 & No \\
		\vspace{-0. cm}& Oberon & Trailing & '' & '' & 2.28 $\pm$ 0.53 & 0.32 $\pm$ 0.08 & Yes \\
          & & Leading & - & - & - & - & No \\
		\hline  
        
        \vspace{-0. cm}4.90 $\micron$ & Ariel & Trailing & 4.898 & 4.845 -- 4.927 & 20.79 $\pm$ 0.61 & 6.98 $\pm$ 0.09 & Yes \\
        Band & & Leading & 4.901 & '' & 6.30 $\pm$ 0.59 & 1.73 $\pm$ 0.09 & Yes \\
		\vspace{-0. cm}& Umbriel & Trailing & 4.900 & '' & 6.05 $\pm$ 0.67 & 2.09 $\pm$ 0.08 & Yes \\
          & & Leading & 4.901 & '' & 1.06 $\pm$ 0.57 & 0.41 $\pm$ 0.08 & No \\
		\vspace{-0. cm}& Titania & Trailing & 4.901 & '' & 3.83 $\pm$ 0.89 & 1.24 $\pm$ 0.09 & Yes \\
          & & Leading & '' & '' & 0.99 $\pm$ 0.71 & 0.37 $\pm$ 0.09 & No \\
		\vspace{-0. cm}& Oberon & Trailing & 4.895 & '' & 2.03 $\pm$ 0.63 & 0.93 $\pm$ 0.08 & Yes \\
          & & Leading & - & - & - & - & No \\  
          \hline
          
        \vspace{-0. cm}5.17 $\micron$ & Ariel & Trailing & 5.171 & 5.115 -- 5.190 & 21.49 $\pm$ 1.66 & 6.36 $\pm$ 0.18 & Yes \\
          Band & & Leading & '' & '' & 5.72 $\pm$ 1.74  & 0.81 $\pm$ 0.21 & Yes \\
        \vspace{-0. cm}& Umbriel & Trailing & '' & '' & 8.58 $\pm$ 2.38 & 2.84 $\pm$ 0.19 & Yes \\
          & & Leading & '' & '' & 2.30 $\pm$ 1.75 & 0.62 $\pm$ 0.20 & No \\
        \vspace{-0. cm}& Titania & Trailing & '' & '' & 4.59 $\pm$ 1.34 & 1.48 $\pm$ 0.19 & Yes \\
          & & Leading & - & - & - & - & No \\
        \vspace{-0. cm}& Oberon & Trailing & 5.171 & 5.115 -- 5.190 & 3.14 $\pm$ 1.23 & 0.94 $\pm$ 0.20 & No \\
          & & Leading & - & - & - & - & No \\          
        \hline            
        
    \label{measurements_pt1} 
	\end{tabular}

\vspace{-0.35 cm}\hspace{-4 cm} \textit{$^a$The 4.20 and 4.25 $\micron$ peaks are convolved, and we report the same area for both features.}
    \end{table}

    \begin{table}[t]
    \centering
	\caption {Measurements of likely carbon oxide spectral features on Ariel and Umbriel.} 
	\vspace{-0.4cm}\begin{tabular}{cccccccc}
		\hline\hline
		\begin{tabular}[c]{@{}l@{}}\hspace{-1 cm}Feature \\ \hspace{-1 cm}Name \end{tabular} & 
		\begin{tabular}[c]{@{}l@{}}\hspace{-1 cm}Satellite \\ \end{tabular} &
		\begin{tabular}[c]{@{}l@{}}\hspace{-1 cm}Hemisphere \end{tabular} &         
		\begin{tabular}[c]{@{}l@{}}\hspace{-1 cm}Feature \\ \hspace{-1 cm}Center ($\micron$) \end{tabular} &  
		\begin{tabular}[c]{@{}l@{}}\hspace{-1 cm}Wavelength \\ \hspace{-1 cm} Range ($\micron$) \end{tabular} & 
		\begin{tabular}[c]{@{}l@{}}\hspace{-1 cm}Spectral \\ \hspace{-1 cm}Contrast ($\%$) \end{tabular} &
		\begin{tabular}[c]{@{}l@{}}\hspace{-1 cm}Band Area \\ \hspace{-1 cm}(10$^-$$^4$ $\micron$) \end{tabular} &		
        \begin{tabular}[c]{@{}l@{}}\hspace{-1 cm}$>$3$\sigma$ Spec. Contrast \\ \hspace{-1 cm}\& Band Area? \end{tabular} \\
		\hline

        \vspace{-0. cm}3.33 $\micron$ & Ariel & Trailing & 3.331 & 3.328 -- 3.357 & 2.73 $\pm$ 0.34 & 3.78 $\pm$ 0.43 & Yes \\
        Band & & Leading &  - & - & - & - & No \\
		\vspace{-0. cm}& Umbriel & Trailing & 3.331 & 3.328 -- 3.357 & 1.50 $\pm$ 0.33 & 1.14 $\pm$ 0.35 & Yes \\
          & & Leading &  - & - & - & - & No \\
        \hline
        
        \vspace{-0. cm}4.30 $\micron$ & Ariel & Trailing & 4.298 & 4.292 -- 4.302 & 2.96 $\pm$ 0.35 & 0.92 $\pm$ 0.26 & Yes \\
        Band & & Leading & '' & '' & 2.75 $\pm$ 0.35 & 1.00 $\pm$ 0.26 & Yes \\
		\vspace{-0. cm}& Umbriel & Trailing & - & - & - & - & No \\
          & & Leading & - & - & - & - & No \\
        \hline

        \vspace{-0. cm}4.38 $\micron$ & Ariel & Trailing &4.378 & 4.365 -- 4.391 & 11.80 $\pm$ 1.10 & 1.76 $\pm$ 0.05 & Yes \\
          Band & & Leading & '' & '' & 8.84 $\pm$ 0.74 & 1.18 $\pm$ 0.05 & Yes \\
		\vspace{-0. cm}& Umbriel & Trailing & 4.379 & 4.367 -- 4.390 & 2.28 $\pm$ 0.51 & 0.27 $\pm$ 0.03 & Yes \\ 
          & & Leading & 4.381 & 4.367 -- 4.390 & 1.55 $\pm$ 0.77 & 0.21 $\pm$ 0.04 & No \\
        \hline
     
		\vspace{-0. cm}4.40 $\micron$ & Ariel & Trailing & 4.403 & 4.393 -- 4.429 & 3.16 $\pm$ 0.51 & 8.06 $\pm$ 0.43 & Yes \\
        Lobe$^a$ & & Leading & '' & '' & 2.25 $\pm$ 0.69 & 2.37 $\pm$ 0.46 & Yes \\
		\vspace{-0. cm}& Umbriel & Trailing & 4.401 & 4.388 -- 4.422 & 5.43 $\pm$ 0.52 & 3.52 $\pm$ 0.36 & Yes \\
          & & Leading & 4.398 & 4.389 -- 4.412 & 3.89 $\pm$ 1.24 & 3.87 $\pm$ 0.38 & Yes \\
        \hline
          
		\vspace{-0. cm}4.41 $\micron$ & Ariel & Trailing & 4.412 & 4.393 -- 4.429 & 5.42 $\pm$ 0.55 & 8.06 $\pm$ 0.43 & Yes \\
        Lobe$^a$ & & Leading & 4.414 & 4.393 -- 4.421 & 2.06 $\pm$ 0.53 & 2.37 $\pm$ 0.46 & Yes \\
		\vspace{-0. cm}& Umbriel & Trailing & 4.416 & 4.388 -- 4.422 & 0.26 $\pm$ 0.45 & 3.52 $\pm$ 0.36 & No \\
          & & Leading & - & - & - & - & No \\
        \hline        

		\vspace{-0. cm}4.47 $\micron$ & Ariel & Trailing & 4.468 & 4.430 -- 4.495 & 6.22 $\pm$ 0.60 & 28.79 $\pm$ 0.66 & Yes \\
        Band & & Leading & 4.463 & '' & 2.25 $\pm$ 0.69 & 9.86 $\pm$ 0.87 & Yes \\
		\vspace{-0. cm}& Umbriel & Trailing & '' & '' & 1.56 $\pm$ 0.58 & 7.77 $\pm$ 0.59 & No \\
          & & Leading & '' & '' & 0.45 $\pm$ 0.71 & 5.67 $\pm$ 0.59 & No \\
         \hline    

        \vspace{-0. cm}4.59 $\micron$ & Ariel & Trailing & 4.591 & 4.552 -- 4.640 & 2.40 $\pm$ 0.43 & 15.14 $\pm$ 0.79 & Yes \\
        Band & & Leading & '' & '' & 1.10 $\pm$ 0.44 & 3.73 $\pm$ 0.89 & No \\
		\vspace{-0. cm}& Umbriel & Trailing & '' & '' & 1.35 $\pm$ 0.67 & 8.20 $\pm$ 0.80 & No \\
          & & Leading & - & - & - & - & No \\
          
         \hline    
        \vspace{-0. cm}4.78 $\micron$ & Ariel & Trailing & 4.781 & 4.764 -- 4.797 & 3.69 $\pm$ 0.88 & 2.54 $\pm$ 0.51 & Yes \\
        Band & & Leading & - & - & - & - & No \\
		\vspace{-0. cm}& Umbriel & Trailing & - & - & - & - & No \\
          & & Leading & - & - & - & - & No \\
		\hline 
        
        \vspace{-0. cm}5.23 $\micron$ & Ariel & Trailing & 5.226 & 5.209 -- 5.237 & 12.36 $\pm$ 1.94 & 1.34 $\pm$ 0.18 & Yes \\
          Band & & Leading & - & - & - & - & No \\          
		\vspace{-0. cm}& Umbriel & Trailing & 5.227 & 5.209 -- 5.237 & 5.19 $\pm$ 1.62 & 0.68 $\pm$ 0.18 & Yes \\          
          & & Leading & - & - & - & - & No \\
        \hline 
    \label{measurements_pt2}    
	\end{tabular}

\vspace{-0.35 cm}\hspace{-4 cm} \textit{$^a$The 4.40 and 4.41 $\micron$ lobes are convolved, and we report the same area for both features.}
    \end{table}

    \begin{table}[t]
    \centering
	\caption {Measurements of H$_2$O ice spectral features on Ariel, Umbriel, Titania, and Oberon.} 
	\vspace{-0.4cm}\begin{tabular}{cccccccc}
		\hline\hline
		\begin{tabular}[c]{@{}l@{}}\hspace{-1 cm}Feature \\ \hspace{-1 cm}Name \end{tabular} & 
		\begin{tabular}[c]{@{}l@{}}\hspace{-1 cm}Satellite \\ \end{tabular} &
		\begin{tabular}[c]{@{}l@{}}\hspace{-1 cm}Hemisphere \end{tabular} &         
		\begin{tabular}[c]{@{}l@{}}\hspace{-1 cm}Feature \\ \hspace{-1 cm}Center ($\micron$) \end{tabular} &  
		\begin{tabular}[c]{@{}l@{}}\hspace{-1 cm}Wavelength \\ \hspace{-1 cm} Range ($\micron$) \end{tabular} & 
		\begin{tabular}[c]{@{}l@{}}\hspace{-1 cm}Spectral \\ \hspace{-1 cm}Contrast ($\%$) \end{tabular} &
		\begin{tabular}[c]{@{}l@{}}\hspace{-1 cm}Band Area \\ \hspace{-1 cm}(10$^-$$^3$ $\micron$) \end{tabular} &		
        \begin{tabular}[c]{@{}l@{}}\hspace{-1 cm}$>$3$\sigma$ Spec. Contrast \\ \hspace{-1 cm}\& Band Area? \end{tabular} \\
		\hline
        \vspace{-0. cm}3.10 $\micron$ & Ariel & Trailing & 3.102 & 3.014 -- 3.168 & 66.00 $\pm$ 1.08 & 38.53 $\pm$ 0.12 & Yes \\
        Peak & & Leading & 3.104 & 3.014 -- 3.168 & 94.30 $\pm$ 1.97 & 55.33 $\pm$ 0.13 & Yes \\
		\vspace{-0. cm}& Umbriel & Trailing & 3.106 & 3.014 -- 3.168 & 82.32 $\pm$ 2.66 & 49.65 $\pm$ 0.11 & Yes \\
          & & Leading & 3.107 & 3.014 -- 3.168 & 96.80 $\pm$ 3.42 & 60.17 $\pm$ 0.12 & Yes \\
		\vspace{-0. cm}& Titania & Trailing & 3.104 & 3.014 -- 3.168 & 134.81 $\pm$ 3.14 & 79.25 $\pm$ 0.09 & Yes \\
          & & Leading & '' & '' & 139.24 $\pm$ 3.34 & 82.32 $\pm$ 0.14 & Yes \\
		\vspace{-0. cm}& Oberon & Trailing & '' & '' & 132.34 $\pm$ 3.37 & 79.41 $\pm$ 0.15 & Yes \\
          & & Leading & '' & '' & 134.47 $\pm$ 3.34 & 80.36 $\pm$ 0.16 & Yes \\
        \hline
        
        \vspace{-0. cm}3.60 $\micron$ & Ariel & Trailing & 3.569 & 3.3 -- 3.9 & 116.80 $\pm$ 0.50 & 424.71 $\pm$ 0.23 & Yes \\
        Cont. Peak & & Leading & 3.564 & 3.3 -- 3.9 & 122.62 $\pm$ 0.52 & 439.51 $\pm$ 0.21 & Yes \\
		\vspace{-0. cm}& Umbriel & Trailing & 3.569 & 3.3 -- 3.9 & 106.22 $\pm$ 0.57 & 394.55 $\pm$ 0.19 & Yes \\
          & & Leading & '' & '' & 109.66 $\pm$ 0.43 & 403.51 $\pm$ 0.19 & Yes \\
		\vspace{-0. cm}& Titania & Trailing & 3.557 & 3.3 -- 3.9 & 138.35 $\pm$ 0.55 & 490.38 $\pm$ 0.17 & Yes \\
          & & Leading & 3.569 & 3.3 -- 3.9 & 140.90 $\pm$ 0.55 & 498.05 $\pm$ 0.22 & Yes \\
		\vspace{-0. cm}& Oberon & Trailing & '' & '' & 126.32 $\pm$ 0.51 & 455.81 $\pm$ 0.23 & Yes \\
          & & Leading & '' & '' & 128.66 $\pm$ 0.56 & 461.86 $\pm$ 0.22 & Yes \\
        \hline
        
        \vspace{-0. cm}4.14 $\micron$ & Ariel$^a$ & Trailing & 4.141 & 4.124 -- 4.156 & 2.60 $\pm$ 0.60 & 0.46 $\pm$ 0.05 & Yes \\
        Band & & Leading & '' & '' & 2.80 $\pm$ 0.84 & 0.50 $\pm$ 0.04 & Yes \\
		\vspace{-0. cm}& Umbriel$^a$ & Trailing & '' & '' & 2.14 $\pm$ 0.67 & 0.34 $\pm$ 0.04 & Yes \\
          & & Leading & '' & '' & 2.13 $\pm$ 0.56 & 0.40 $\pm$ 0.04 & Yes \\
		\vspace{-0. cm}& Titania & Trailing & '' & '' & 3.03 $\pm$ 0.11 & 0.42 $\pm$ 0.04 & Yes \\
          & & Leading & '' & '' & 3.71 $\pm$ 0.63 & 0.59 $\pm$ 0.04 & Yes \\
		\vspace{-0. cm}& Oberon & Trailing & '' & '' & 2.52 $\pm$ 0.66 & 0.36 $\pm$ 0.04 & Yes \\
          & & Leading & '' & '' & 2.80 $\pm$ 0.63 & 0.51 $\pm$ 0.04 & Yes \\
          \hline
    \label{measurements_pt3} 
	\end{tabular}

\vspace{-0.35 cm} \hspace{-0.8 cm} \textit{$^a$The 4.14 $\micron$ bands for the inner moons Ariel and Umbriel are likely contaminated by contributions from nearby CO$_2$} 

\hspace{-0.3 cm} \textit{features. For completeness, we include measurements of the 4.14 $\micron$ band on Ariel and Umbriel, with the caveat that they}

\hspace{-14.4 cm} \textit{are probably unreliable.}
    \end{table}

    \newpage
    
	\subsection{Comparison to Laboratory Spectra} 
             
\begin{figure}[h!]
         \centering
			\includegraphics[scale=0.85]{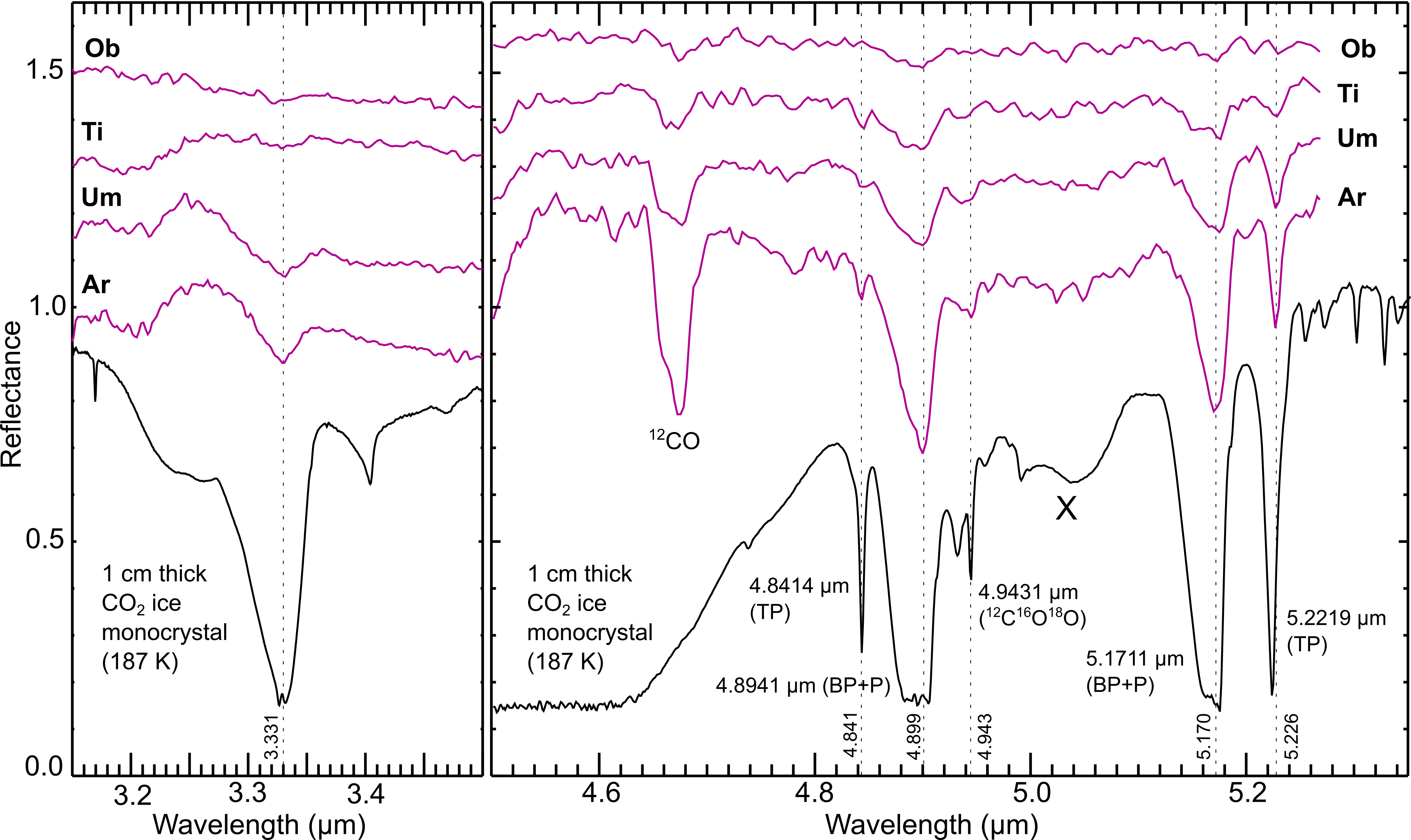}
		 \centering
			\caption{\textit{Left: Trailing/leading spectral ratios for Ariel (Ar), Umbriel (Um), Titania (Ti), and Oberon (Ob) compared to a laboratory spectrum of a $\sim$1 cm thick monocrystal of CO$_2$ ice (187 K) between 3.15 and 3.5 $\micron$, offset vertically for clarity. The central wavelength ($\micron$) of the 3.331 $\micron$ feature (secure detection; Table \ref{vibrational_modes}) identified on Ariel and Umbriel (listed vertically along the dotted line) is in close agreement with the center of a strong absorption feature in the laboratory spectrum. Right: The same spectral ratios showing spectral features near 4.899, 5.171, and 5.226 $\micron$ (secure detections; Table \ref{vibrational_modes})} that align well with the positions of triphonon (TP) and biphonon plus phonon (BP+P) modes identified in crystalline CO$_2$ ice \citep{bini1991triphonons}, although the 5.226 $\micron$ feature observed on the Uranian moons appears to be shifted to slightly longer wavelengths compared to the TP mode expressed by the laboratory data ($\sim$5.222 $\micron$). The laboratory spectrum is saturated between $\sim$3.95 and 4.65 $\micron$ due to strong absorption by the CO$_2$ $\nu$$_3$ mode. The Uranian moon spectral ratios exhibit an additional feature near 4.67 $\micron$ (Table \ref{measurements_pt1}) that is absent from crystalline CO$_2$ ice and results from $^{12}$CO ice. Other features exhibited by the Uranian moons (especially Ariel) near 4.84 and 4.94 $\micron$ aligns well with a TP mode in $^{12}$CO$_2$ and a $^{12}$C$^{16}$O$^{18}$O feature, respectively (tentative detections; Table \ref{vibrational_modes}). Additional `continuum' structure in the laboratory spectrum may result from unattributed modes in crystalline CO$_2$ ice (example marked with an `X').}
        \label{CO2_phonon_modes} 
		\end{figure}

\begin{SCfigure}[1.1][h!]
         \centering
			\includegraphics[width=0.57\textwidth]{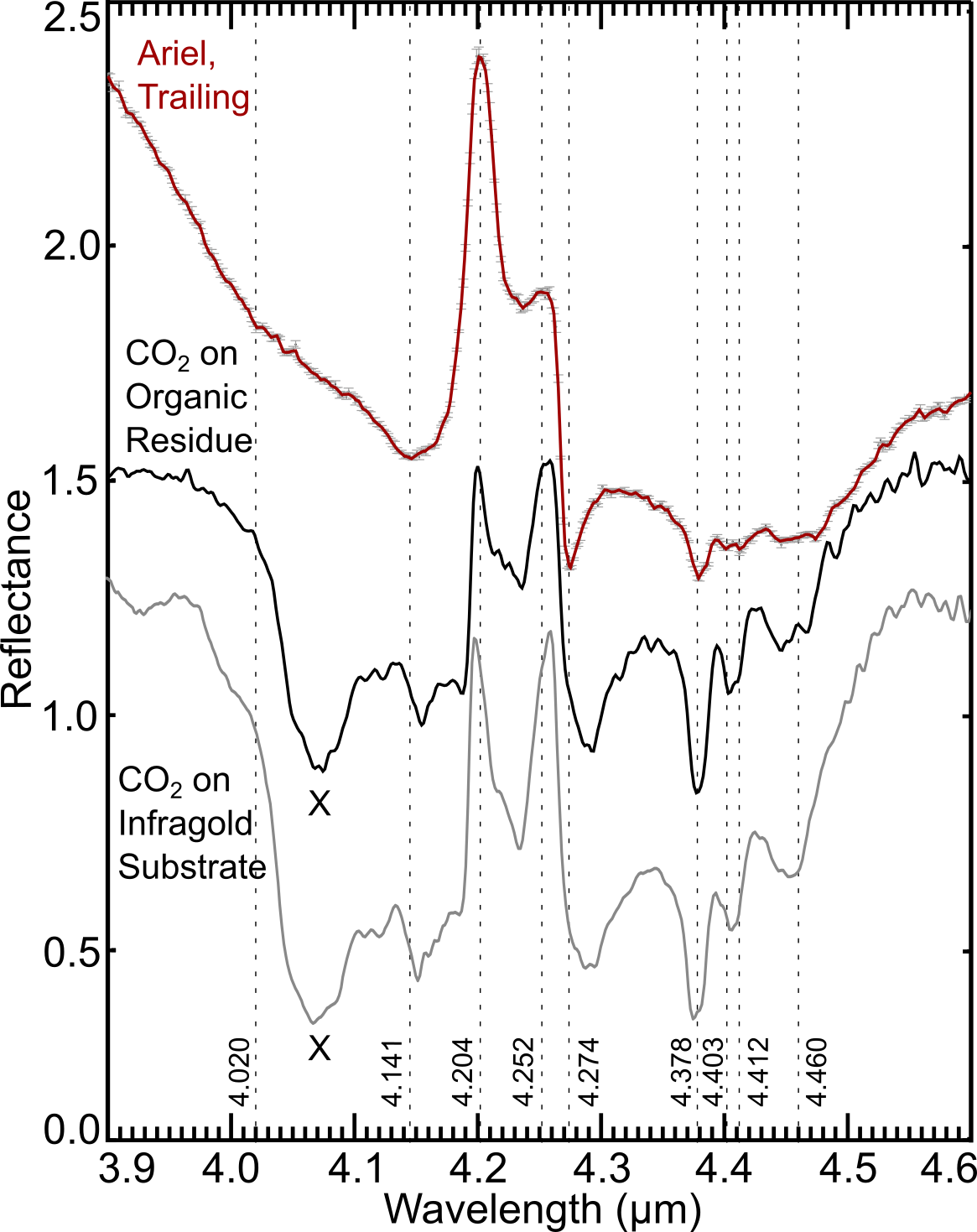}
		 \centering
			\caption{\textit{G395M data of Ariel's trailing hemisphere compared to laboratory reflectance data of a $\sim$10 $\micron$ thick layer of CO$_2$ ice deposited on an irradiated organic residue and $\sim$15 $\micron$ thick layer of CO$_2$ ice deposited directly onto an Infragold substrate (both warmed up to 90 K), arbitrarily scaled and offset vertically for clarity. The central wavelengths ($\micron$) for features with secure, probable, and tentative detections (Table \ref{vibrational_modes}) in the Ariel spectrum are listed vertically along each dotted line (measurements summarized in Tables \ref{measurements_pt1}, \ref{measurements_pt2}, and \ref{measurements_pt3}). A possible Christiansen feature expressed by the laboratory spectra (marked with an `X') is notably absent from G395M data of Ariel and the other large moons.}}
        \label{CO2_4.0-4.6} 
		\end{SCfigure}

\begin{SCfigure}[1.1][h!]
			\includegraphics[width=0.5\textwidth]{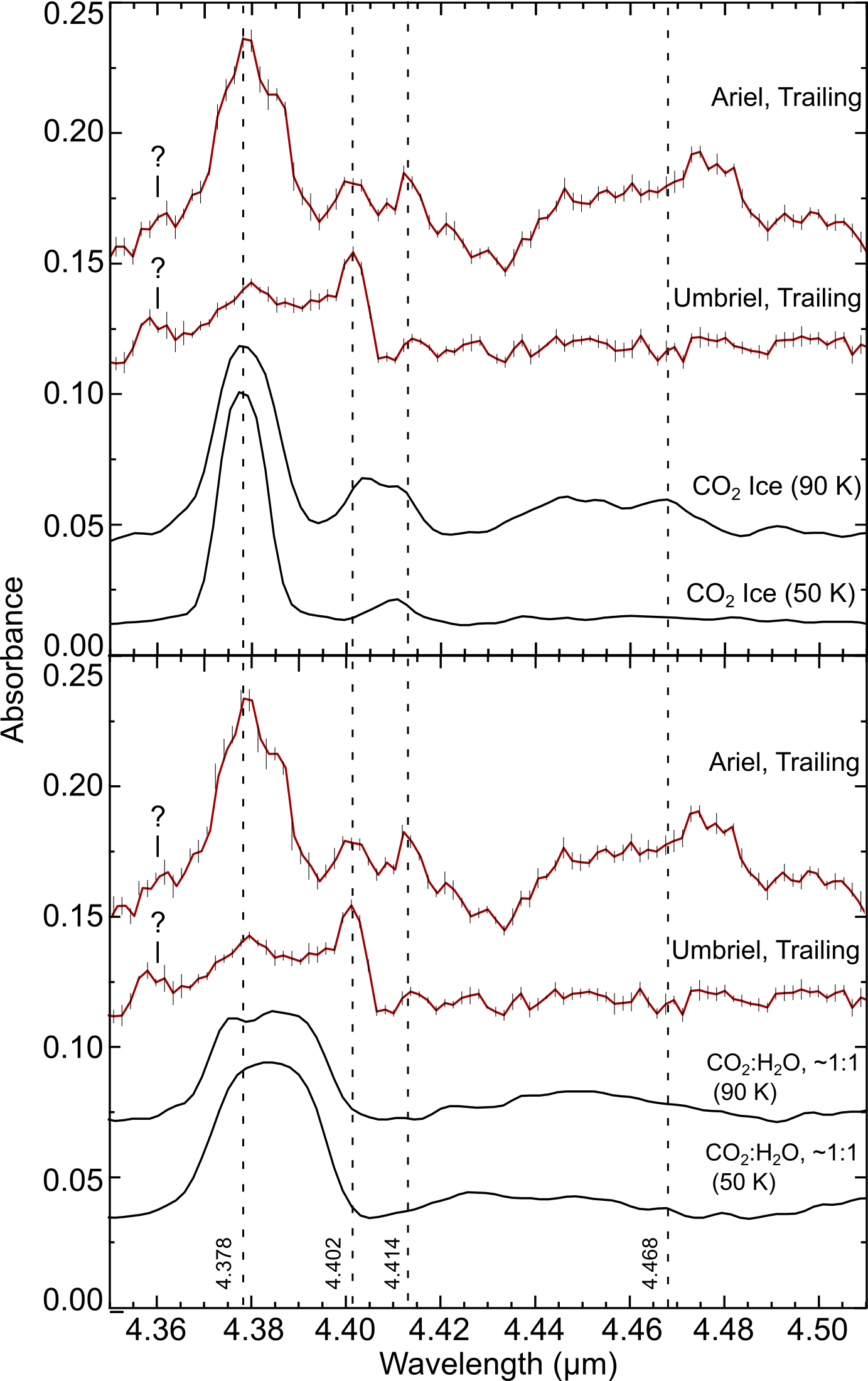}
		    \caption{\textit{Top: Absorbance spectra (\textit{--log(reflectance)}) of a thick CO$_2$ ice layer deposited at 50 K and warmed up to 90 K on an organic residue, measured using INGMAR \citep[e.g.,][]{henault2025}. These laboratory spectra are compared to G395M reflectance data of Ariel trailing and Umbriel trailing, converted to absorbance (\textit{--log(reflectance)}) and offset vertically for clarity. Bottom: CO$_2$ ice co-deposited with H$_2$O ice at 50 K and then warmed up to 90 K, compared to the same Ariel trailing and Umbriel trailing data in the top plot. Dashed lines highlight spectral features identified on Ariel, some of which are also seen on Umbriel, with the approximate central wavelength listed vertically along the lines. The 4.38, 4.40, and 4.41 $\micron$ features identified on Ariel match similar spectral features in the CO$_2$ ice layer deposited on an organic residue, in particular at 90 K, comparable to the estimated peak temperatures on the Uranian moons \citep{hanel1986infrared, sori2017wunda}. The match to the Umbriel data, however, is less certain. Furthermore, additional structure in the CO$_2$ ice laboratory spectrum (90 K) provides a rough match to the short wavelength side of Ariel's $\sim$4.47 $\micron$ feature, hinting that CO$_2$ might contribute to this feature, along with other compounds. In contrast, the CO$_2$:H$_2$O mixtures do not provide reasonable matches to either Ariel's or Umbriel's 4.38 $\micron$ features and do not reproduce the features near 4.4, 4.41, or 4.47 $\micron$. None of the laboratory spectra are able to reproduce a small feature near 4.36 $\micron$ detected in the Ariel and Umbriel data (labeled with question marks).}}
            \label{13CO2_4.34-4.52µm} 
        \end{SCfigure}

\begin{SCfigure}[1.1][h!]
			\includegraphics[width=0.5\textwidth]{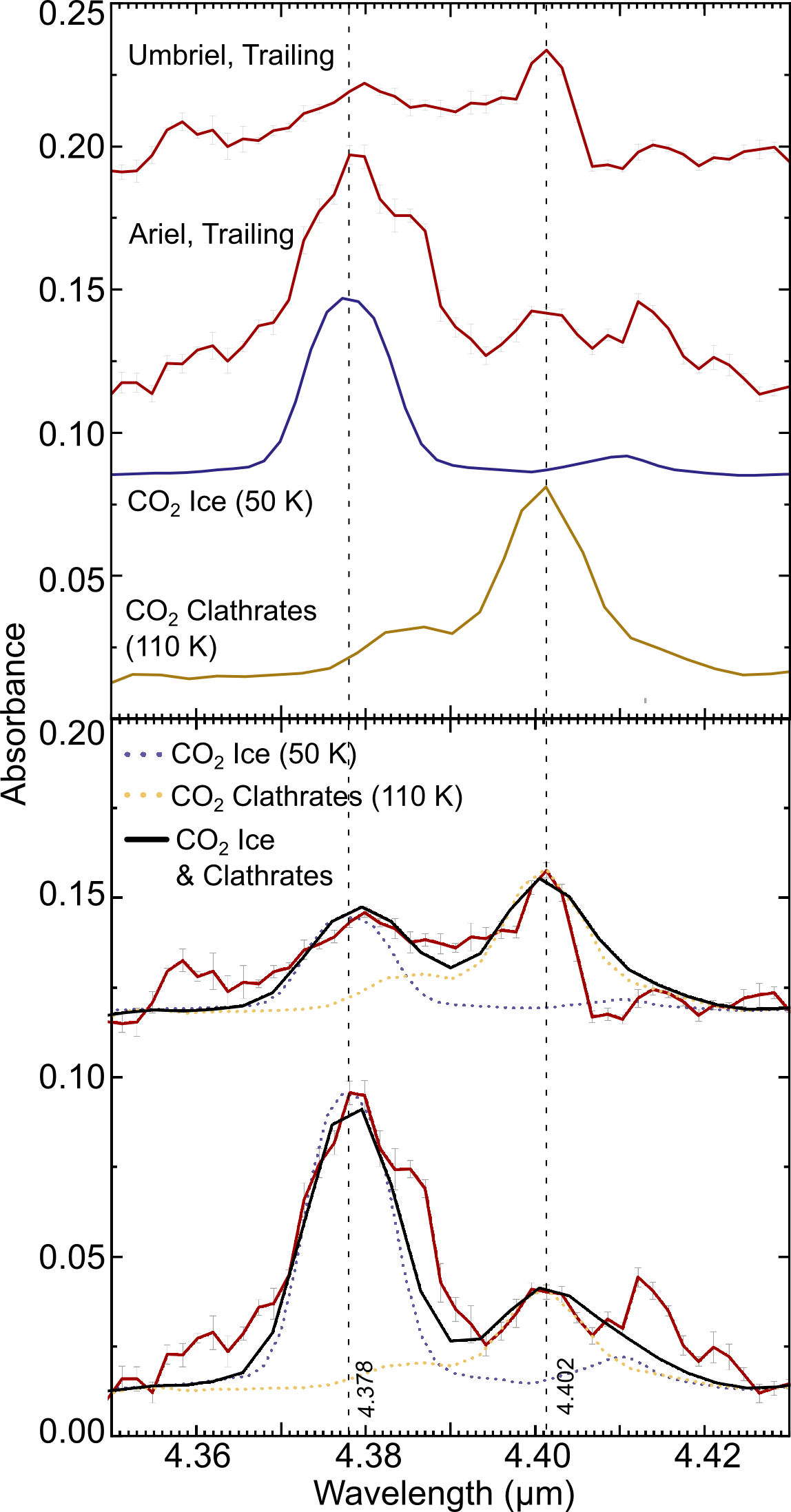}
			\caption{\textit{Top: Absorbance spectra (\textit{--log(reflectance)}) of CO$_2$ ice (50 K) measured using INGMAR \citep[e.g.,][]{henault2025} and CO$_2$ clathrates (110 K) \citep{oancea2012CO2clathrates}. These laboratory spectra are compared to G395M reflectance data of Ariel trailing and Umbriel trailing, converted to absorbance (\textit{--log(reflectance)}) and offset vertically for clarity. Bottom: CO$_2$ ice and CO$_2$ clathrates scaled to fit the G395M data and linear combinations of these scaled laboratory spectra fit to the G395M data, with Ariel and Umbriel offset vertically for clarity. These scaled spectra of CO$_2$ ice and CO$_2$ clathrates provide reasonable matches to the 4.38 $\micron$ and 4.4 $\micron$ features, respectively. Additional features near 4.36 $\micron$ and 4.41 $\micron$ are not well matched by these laboratory species. These scaled spectral data offer useful but non-unique solutions, and other scaled combinations warrant study in future work.}}
            \label{13CO2_4.34-4.44µm} 
		\end{SCfigure}

\begin{SCfigure}[1.1][h!]
			\includegraphics[width=0.5\textwidth]{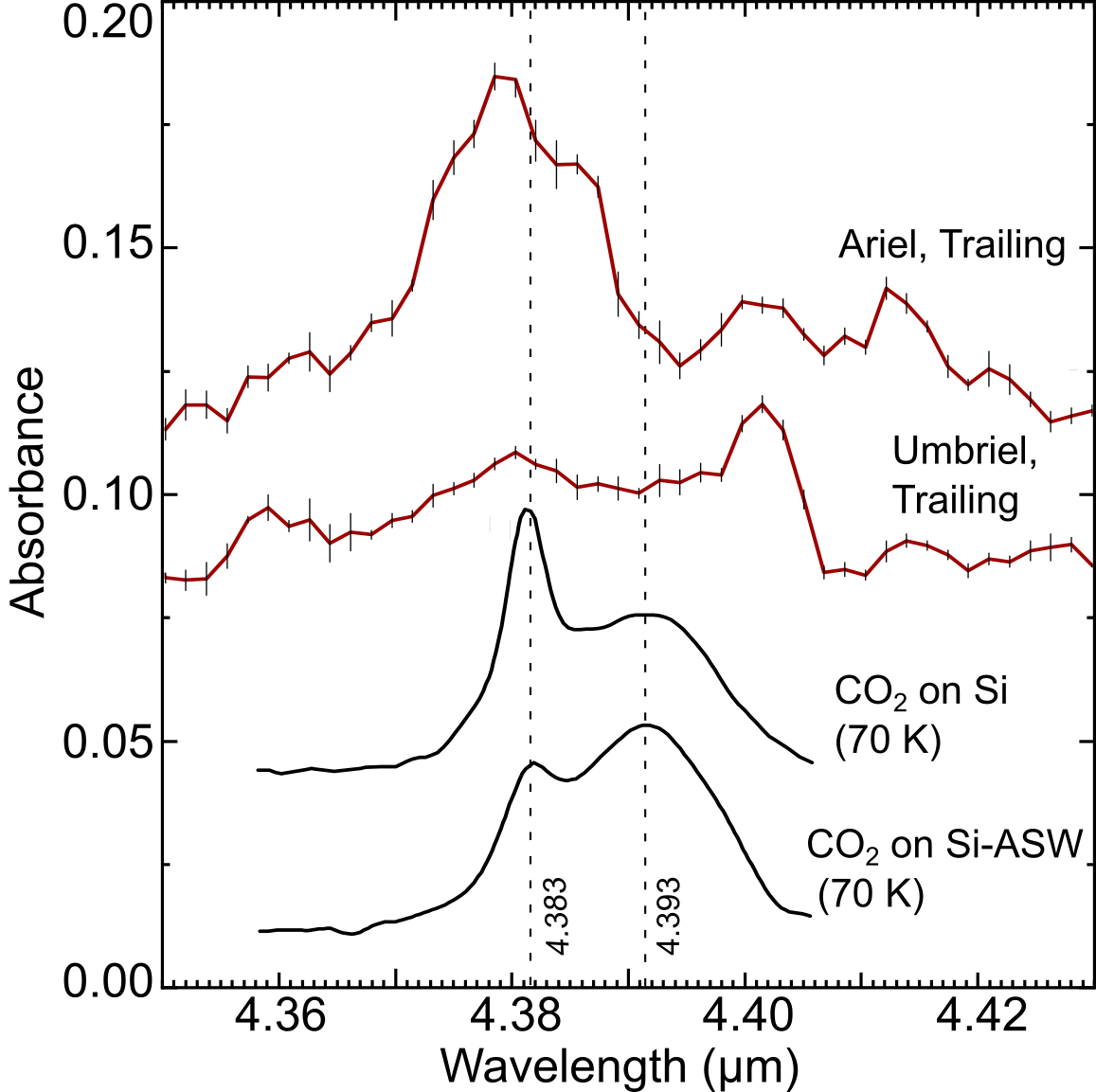}
			\caption{\textit{Laboratory spectra (modified from \citealt{suhasaria2025CO2dust}) of CO$_2$ ice condensed on a silicate-rich, meteorite simulant layer (70 K) and amorphous H$_2$O ice (`ASW') deposited on a meteorite simulant layer (70 K). These laboratory spectra are compared to G395M reflectance data of Ariel trailing and Umbriel trailing, converted to absorbance (\textit{--log(reflectance)}) and offset vertically for clarity. Dashed lines highlight spectral features identified in the laboratory data near 4.383 and 4.393 $\micron$. The 4.383 $\micron$ peak is offset from the Ariel and Umbriel data, but the 4.393 $\micron$ peak provides a potentially useful match to Umbriel's broad $^{13}$CO$_2$ feature that warrants study in future work.}}
            \label{13CO2_Suhasaria} 
		\end{SCfigure}

        After H$_2$O ice, the spectral properties of Ariel and the other large Uranian moons are largely defined by CO$_2$, dominated by a few prominent features associated with its $\nu$$_3$ mode and complemented by a rich variety of more subtle CO$_2$ features. To help isolate the signature of CO$_2$ features outside the core of the CO$_2$ $\nu$$_3$ mode (4.2 -- 4.3 $\micron$), we generated spectral ratios by dividing the trailing hemisphere spectrum by the leading hemisphere spectrum for each moon (Figure \ref{spectral_ratios}). To better understand the nature of combination and overtone modes in CO$_2$ ice between 3.3 -- 3.4 $\micron$ and between 4.8 -- 5.3 $\micron$, we compared the spectral ratios for all four moons to a spectrum of a $\sim$1 cm thick monocrystal of CO$_2$ ice, generated using procedures described in section 2.2 (Figure \ref{CO2_phonon_modes}). To investigate the nature of isotopologues and help determine whether other carbon oxides are present, we compared the G395M data of Ariel and Umbriel, the two moons with the strongest CO$_2$ features, to laboratory absorbance and reflectance spectra of crystalline CO$_2$ ice and CO$_2$ clathrates (Figures \ref{CO2_4.0-4.6}, \ref{13CO2_4.34-4.52µm}, \ref{13CO2_4.34-4.44µm}, and \ref{13CO2_Suhasaria}). 
         
        \textit{Trailing/Leading spectral ratios compared to thick cell CO$_2$ ice sample.} Optically thick CO$_2$ ice substrates measured in the laboratory express a variety of weak absorption features between 4.8 and 5.25 $\micron$, resulting from the $\nu_1$+$\nu_2$ and 3$\nu_2$ vibration modes in Fermi resonance \citep{dows1973CO2phonons,bini1991triphonons,quirico1997CO2SSHADE,quirico1997near}. Several bands are observed in this wavelength region, including a broad continuum of absorption due to the overlap of the $\nu_2(\vec{k}_1)$+$\nu_2(\vec{k}_2)$+$\nu_2(\vec{k}_3)$ ($\vec{k}_1$+$\vec{k}_2$+$\vec{k}_3$=$\vec{0}$) and the $\nu_1(\vec{k}_1)$+$\nu_2(\vec{k}_2)$ ($\vec{k}_1$+$\vec{k}_2$=$\vec{0}$) continua. Identified features include two sharp triphonon (TP) modes at 4.842 and 5.222 $\micron$, as well as two biphonon+phonon (BP+P) bands near 4.894 and 5.171 $\micron$ that are both saturated in the presented laboratory spectrum (Figure \ref{CO2_phonon_modes}). The laboratory data also show two weak and narrow features near 4.902 $\micron$ (merged in the shoulder of the BP+P band) and 4.943 $\micron$, which are due to the $^{13}$CO$_2$ and $^{12}$C$^{16}$O$^{18}$O isotopologues, respectively. In particular for Ariel and Umbriel, the G395M data and spectral ratios show features that are well matched by the TP and BP+P modes (secure detections), along with the presence of a weaker feature near 4.94 $\micron$ (tentative detection) that may result from $^{12}$C$^{16}$O$^{18}$O ice (Table \ref{vibrational_modes}). The expression of these TP and BP+P features very likely confirms that crystalline CO$_2$ ice is present on the surface of Ariel and the other moons, supporting earlier ground-based studies that made similar assertions based on the presence of the CO$_2$ triplet band \citep[e.g.,][]{grundy2003discovery}.  Although these moons' 4.90 $\micron$ bands are likely dominated by CO$_2$ ice, a 4.89 $\micron$ feature exhibited by CO$_3$, formed in irradiated CO$_2$ substrates \citep{moll1966CO3,raut2013CO3}, cannot be ruled out as a minor contributor to the short wavelength wing of the 4.90 $\micron$ bands \citep{cartwright2024ArielJWST}. 

        We note additional, subtle structure between $\sim$5 and 5.1 $\micron$ in the G395M data, especially of Ariel, roughly centered near 5.02 $\micron$ (tentative detection; Table \ref{vibrational_modes}). The CO$_2$ monocrystal expresses several other weak absorption features within this same wavelength range that still lack attribution to specific vibrational modes (example labeled with an `X' in Figure \ref{CO2_phonon_modes}). Thus, perhaps the subtle structure between 5 and 5.1 $\micron$ results from weak and unattributed CO$_2$ features. Alternatively, species such as dicarbon monoxide (C$_2$O), a free radical formed via irradiation of carbon suboxide (C$_3$O$_2$) and CO ices at cryogenic temperatures express features centered near 5.02 $\micron$ \citep[e.g.,][]{jacox1965C2O,bennett2010CO2CO3}.       
                 
        Similar to the TP and BP+P modes, the 3.33 $\micron$ band is only observed in environments dominated by crystalline CO$_2$ ice, such as CO$_2$ frost clouds on Mars \citep[e.g.,][]{bell1996CO2clouds}. The vibrational modes that contribute to this band in solid-state CO$_2$ are not well characterized, but a $\sim$3.3 $\micron$ feature observed in experiments involving CO$_2$ \textit{gas} has been attributed to a $\nu$$_2$+$\nu$$_3$ `forbidden' transition mode, stemming from a magnetic dipole (M1) \citep{kazakov2021CO2gasM1}. Future analytical work, and perhaps additional experiments, are required to determine the modes contributing to the 3.33 $\micron$ feature expressed by CO$_2$ ice.

         \textit{Comparison to substrates mantled by CO$_2$ ice and CO$_2$:H$_2$O mixtures.} Ground-based observations indicate that the spectral properties of the large Uranian moons are consistent with H$_2$O ice well-mixed with dark and reddish material(s), capped by isolated deposits of crystalline CO$_2$ ice that likely include minor amounts of other carbon oxides \citep[e.g.,][]{grundy2006distributions, cartwright2022ArielCO2}. To better understand the spectral properties of CO$_2$ mantling these moons' regoliths, we compared the G395M data of Ariel's trailing hemisphere to laboratory spectra of stratified and textured layers, including a $\sim$10 $\micron$ thick layer of CO$_2$ ice condensed onto a $\sim$1 $\micron$ thick, proton-irradiated organic residue and a $\sim$15 $\micron$ thick layer of CO$_2$ ice condensed directly onto the Infragold substrate (Figure \ref{CO2_4.0-4.6}). The laboratory spectra show a prominent, double-lobed scattering peak, with wavelength positions consistent with Ariel's double-lobed scattering peaks, centered near 4.20 and 4.25 $\micron$, demonstrating that layering and surface roughness can enhance CO$_2$ scattering peaks. Although the overall shape and intensity of Ariel's peaks are not well matched by the laboratory data, they demonstrate that CO$_2$ ice veneers capping rough(er) substrates can crudely replicate the scattering peaks. Similarly, synthetic spectra of granular CO$_2$ ice (10 $\micron$ diameter grains; optical constants reported in \citealt{gerakines2020CO2}), are also able to crudely match Ariel's 4.20 $\micron$ and 4.25 $\micron$ peaks \citep{cartwright2024ArielJWST}, highlighting the importance of scattering between grains in a regolith for reproducing these features.
         
         The laboratory spectrum also shows features centered near 4.38, 4.40, 4.41, and 4.46 $\micron$ that have central wavelengths that align well with features in the Ariel spectrum, suggesting that relatively pure CO$_2$ ice might explain much of the spectral structure observed in the G395M data. Indeed, earlier laboratory experiments showed that thick CO$_2$ ice substrates can express notable features at these wavelengths (see Fig. 2 in \citealt{dows1973CO2phonons}). Comparing Ariel's 4.27 $\micron$ feature to the laboratory data, however, reveals a notable disparity with the CO$_2$ $\nu$$_3$ mode shifted to 4.29 $\micron$ in the laboratory spectrum. Similarly, a prominent $\sim$4.07 $\micron$ `absorption' feature in the laboratory data is not observed in the Ariel data. This 4.07 $\micron$ feature is generally attributed to a Christiansen effect in CO$_2$ ice \citep{dePra2025JWSTTNOs}, where the real index \textit{n} approaches unity, leading to reduced internal scattering and lower reflectance, thereby mimicking absorption, even at wavelengths where the extinction coefficient, \textit{k}, is small \citep{hapke2012theory}. However, this feature is significantly weakened in samples where CO$_2$ is embedded in other materials with higher refractive indices \citep{dePra2025JWSTTNOs}. The absence of this 4.07 $\micron$ feature therefore hints that some fraction of CO$_2$ on the Uranian moons is well mixed or perhaps trapped in another compound like H$_2$O ice, hypothetically supporting the presence of CO$_2$ clathrates. 

        \begin{SCfigure}[1.1][h!]
			\includegraphics[width=0.47\textwidth]{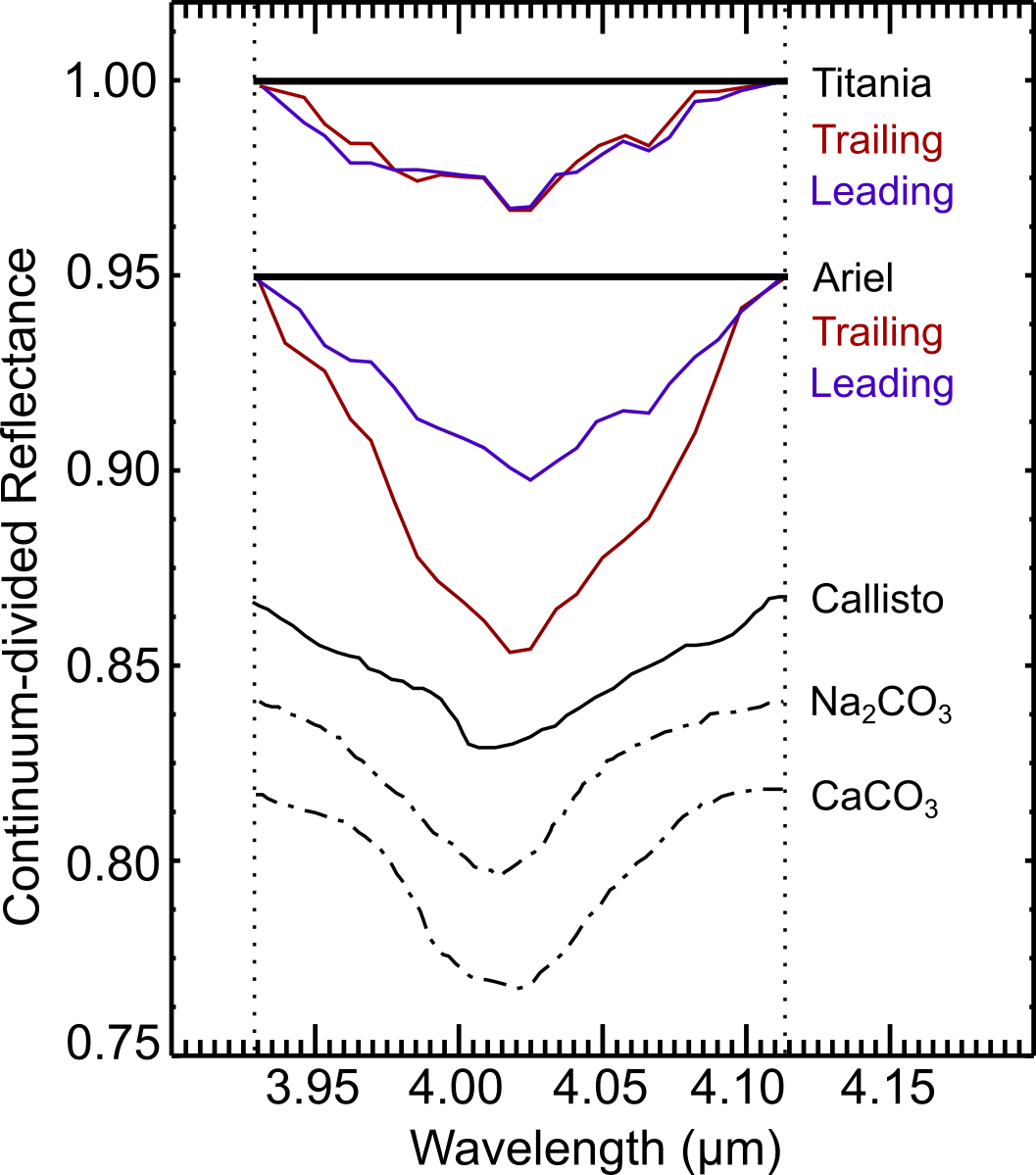}
			\caption{\textit{Continuum-divided reflectance data showing 4 $\micron$ bands identified in the leading and trailing hemisphere data collected at Ariel and Titania (secure detections; Table \ref{vibrational_modes}), IRTF/SpeX data of Callisto's leading hemisphere (originally reported in \citealt{cartwright2020CallistoS}), and laboratory data of anhydrous sodium carbonate and calcium carbonate \citep{nyquist1997handbook}, uncertainties removed and offset vertically for clarity. The 4 $\micron$ features identified in the G395M data and reported in Table \ref{measurements_pt1} (3.96 -- 4.11 $\micron$) are slightly narrower than the feature identified on Callisto (3.92 -- 4.11 $\micron$). Consequently, we re-measured the 4 $\micron$ features in the G395M data of Titania and Ariel using the same wavelength range as Callisto's 4 $\micron$ band, thereby enabling a more suitable comparison between these three moons.}}
            \label{cont-div_4um} 
		\end{SCfigure}

         \textit{Potential band-splitting in $^{13}$CO$_2$ ice.} Between 4.36 and 4.42 $\micron$, the G395M data collected over Ariel's and Umbriel's trailing hemispheres exhibit a multi-lobe structure that has been observed previously in thick CO$_2$ ice deposits (as shown in Fig. 2 of \citealt{dows1973CO2phonons}), as well as CO$_2$ ice mixed with CO$_2$ clathrates \citep{dartois2009CO2clathrates, oancea2012CO2clathrates}.  Other studies have revealed multi-lobe structure in $^{13}$CO$_2$ features between 4.36 and 4.40 $\micron$ in dense clouds enveloping young stars observed by JWST/NIRSpec, possibly resulting from CO$_2$ mixed with H$_2$O ice and/or CH$_3$OH \citep[e.g.,][]{brunken2024JprotostellarJWST}, demonstrating that band-splitting of CO$_2$ features can occur in astrophysical environments. Here, we investigate possible band-splitting in the $^{13}$CO$_2$ features identified in G395M reflectance data of Ariel and Umbriel by first converting these spectra to absorbance (\textit{--log(reflectance)}). We then subtracted off the continua between 4.3 and 4.5 $\micron$, and compared the resulting continua-subtracted G395M data to laboratory data of CO$_2$ co-condensed with H$_2$O ice, CO$_2$ condensed onto a thin organic residue, CO$_2$ condensed on an amorphous MgFeSiO$_4$ meteorite simulant, and a linear mixture of a CO$_2$ ice spectrum combined with a CO$_2$ clathrate spectrum (Figures \ref{13CO2_4.34-4.52µm}, \ref{13CO2_4.34-4.44µm}, and \ref{13CO2_Suhasaria}). The rest of this section focuses on our analysis of these absorbance-converted G395M data of Ariel and Umbriel. This analysis is provided as a starting point to motivate future studies that more rigorously investigate the nature of CO$_2$ on the Uranian moons.       

         The laboratory spectrum of CO$_2$ condensed on an organic residue (warmed to 90 K), expresses additional structure between 4.40 and 4.43 $\micron$, that approximates the central wavelengths of Ariel's 4.40 and 4.41 $\micron$ features (Figure \ref{13CO2_4.34-4.52µm}). This laboratory spectrum also exhibits a broad absorption band between 4.43 and 4.48 $\micron$ that is comparable to the wavelength range of Ariel's 4.47 $\micron$ feature. Similar components may contribute to the spectral structure observed in the G395M data of Umbriel near 4.4 $\micron$, although the Umbriel data do not display 4.41 and 4.47 $\micron$ features, making the match less compelling than for Ariel. The laboratory spectra of CO$_2$ co-condensed with H$_2$O ice does not express a multi-lobe morphology and has a notably broader 4.38 $\micron$ feature, stretching between $\sim$4.37 and 4.40 $\micron$, representing a poorer match to the narrower features expressed by the G395M data of both Ariel and Umbriel (Figure \ref{13CO2_4.34-4.52µm}). An additional 4.36 $\micron$ sideband expressed by the G395M data of these two moons (labeled with a `?' in Figure \ref{13CO2_4.34-4.52µm}) is not observed in any of the laboratory data we analyzed. Determining whether the 4.36 $\micron$ feature results from an unidentified compound or combination of compounds is left for future work. 
         
         Next, we compared the G395M data to a linear mixture of CO$_2$ ice (50 K) condensed onto an organic residue and CO$_2$ clathrates (110 K; \citealt{oancea2012CO2clathrates}), scaled to the Ariel spectrum (0.075 and 0.59, respectively) and Umbriel spectrum (0.026 and 0.85, respectively). This comparison shows that these two components can approximate the shape of the multi-lobe structure expressed by these moons, except for the previously mentioned 4.36 $\micron$ sidebands (Figure \ref{13CO2_4.34-4.44µm}). These scaled laboratory spectra provide a non-unique solution and other scaled combinations may provide reasonable matches. Nonetheless, this comparison shows that mixtures of CO$_2$ ice and CO$_2$ clathrates can provide reasonable matches to the spectral properties of these moons. The laboratory spectra of CO$_2$ condensed on a layer of amorphous MgFeSiO$_4$, and the same amorphous silicate layer capped by an additional layer of amorphous H$_2$O ice \citep{suhasaria2025CO2dust}, exhibit double-lobed $^{13}$CO$_2$ features, centered near 4.383 and 4.393 $\micron$ (Figure \ref{13CO2_Suhasaria}). These laboratory spectra do not exhibit 4.41 $\micron$ features, unlike the Ariel G395M data. However, these silicate-CO$_2$ stratified substrates show a band of absorption between 4.37 and 4.40 $\micron$, appearing to provide a useful match to the G395M data of Umbriel, which exhibits a stronger absorption profile near 4.39 $\micron$ compared to Ariel. 
           
         These comparisons highlight that the finer scale structure of Ariel's $^{13}$CO$_2$ $\nu$$_3$ mode is reasonably approximated by CO$_2$ ice (90 K) condensed on an irradiated organic residue, which may also contribute to its broad 4.47 $\micron$ feature. In this scenario, Ariel's 4.41 $\micron$ feature could include contributions from $^{13}$C$^{16}$O$^{18}$O \citep{loeffler2005CO, bennett2010CO2CO3}. Additionally, Ariel's broad 4.47 $\micron$ feature (spanning $\sim$4.44 to 4.49 $\micron$) could include contributions from $^{13}$C$^{18}$O$_2$ near 4.45 $\micron$ \citep{loeffler2005CO}. The relative strength of Ariel's 4.47 $\micron$ band compared to the laboratory data (Figure \ref{13CO2_4.34-4.52µm}) suggests that other, non-CO$_2$ compounds may contribute as well, such as carbon-chain oxides and nitriles.  CO$_2$ ice (50 K) mixed with CO$_2$ clathrates (110 K) can also reproduce the multi-lobe structure expressed in the G395M data of Ariel and Umbriel between 4.36 and 4.42 $\micron$. The structure of Umbriel's $^{13}$CO$_2$ $\nu$$_3$ mode is not well matched by any one of the laboratory spectra inspected here, but could be better matched by a combination of different laboratory spectra, including CO$_2$ co-condensed with H$_2$O or CO$_2$ condensed on amorphous silicates. 
           
         \textit{Does CO$_2$ contribute to the 4.02 $\micron$ band?} Similar to the measurements of CO$_2$ and CO reported here, the 4.02 $\micron$ feature (secure detection; Table \ref{vibrational_modes}) is strongest on Ariel, in particular on its trailing hemisphere. However, the next strongest 4.02 $\micron$ bands are on Titania, followed by Oberon, and the weakest 4.02 $\micron$ bands are on Umbriel, unlike the trends in CO$_2$ and CO (Table \ref{measurements_pt1}), indicating that this feature likely results from other species. Furthermore, none of the CO$_2$-dominated laboratory spectra presented here, or synthetic spectra of CO$_2$ ice presented previously \citep{cartwright2024ArielJWST}, show features consistent with the 4.02 $\micron$ band. Although the continuum of H$_2$O ice exhibits a pronounced slope change in this wavelength region that could hypothetically contribute to the 4.02 $\micron$ band, the strongest H$_2$O and HDO ice features we measured are on Titania, not Ariel, inconsistent with the reported 4.02 $\micron$ band strengths. 
         
         It therefore seems more likely that another component dominates contributions to this feature, such as carbonate minerals, as suggested by a prior study \citep{cartwright2024ArielJWST}. To investigate this possibility further, we compared continuum-divided 4.02 $\micron$ bands of Ariel and Titania to continuum divided 4 $\micron$ features exhibited by laboratory spectra of anhydrous sodium carbonate (Na$_2$CO$_3$) and calcium carbonate (CaCO$_3$), as well as a 4 $\micron$ feature expressed by Jupiter's moon Callisto (Figure \ref{cont-div_4um}), which has been attributed to carbonates \citep{johnson2004radiation,  cartwright2024CallistoJWST}. The laboratory spectra of carbonates and Callisto's 4 $\micron$ feature share similar central wavelengths and morphologies to the 4.02 $\micron$ band on the Uranian moons, providing another piece of evidence supporting the presence of carbonates. Of note, we see no evidence for a 3.4 $\micron$ band complex associated with carbonates \citep{hexter1958CO3}, possibly because it fully overlaps the long wavelength wing of the broad 3 $\micron$ H$_2$O ice band \citep[e.g.,][]{mastrapa2009H2Oopcon}.
         
         Alternatively, the 4.02 $\micron$ band could result from a 4.07 $\micron$ feature in CO$_2$ ice (marked with an `X' in Figure \ref{CO2_4.0-4.6}),  generally attributed to a Christiansen effect where the real part of the refractive index for CO$_2$ drops below unity, reducing internal scattering and limiting reflectance, thereby mimicking the spectral profile of a strong absorption band \citep[e.g.,][]{dePra2025JWSTTNOs}. In this scenario, some currently unidentified process would have to account for the notable difference between the position of this purported Christiansen feature and the 4.02 $\micron$ band ($\sim$0.05 $\micron$ wavelength shift). On balance, carbonate minerals are the most useful match in terms of central wavelength and band morphology, and we favor this interpretation for the origin of the 4 $\micron$ feature on the Uranian moons (Table \ref{vibrational_modes}). Future laboratory experiments involving crystalline CO$_2$ ice and other carbon oxides under conditions relevant to the Uranus system are likely needed to determine whether carbonates or some other combination of species can replicate the morphology of the 4.02 $\micron$ band.

\section{Discussion} 
        
        \subsection{Distribution and spectral signature of CO$_2$}

\begin{figure}[h!]
         \centering
			\includegraphics[scale=0.85]{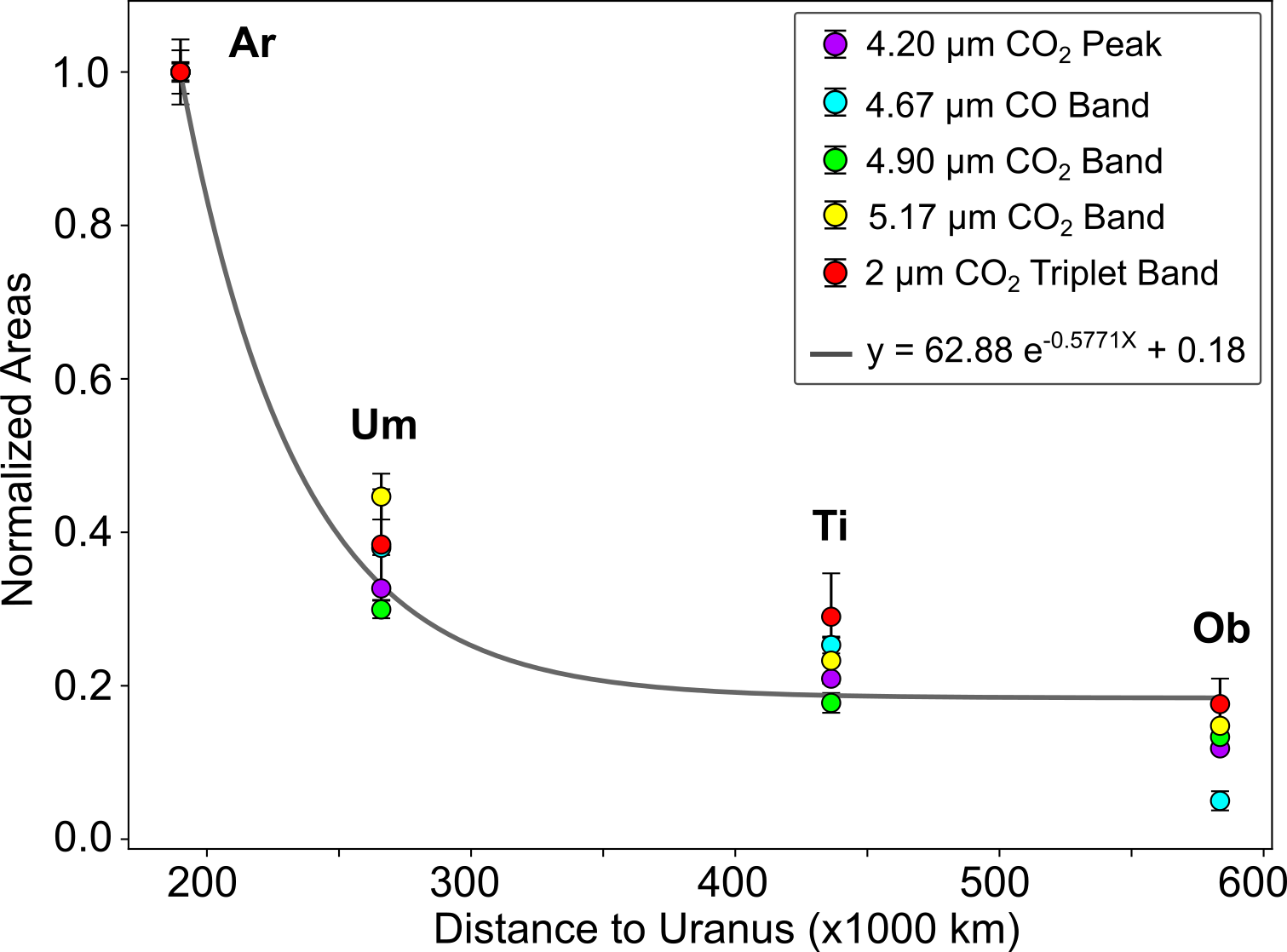}
		 \centering
			\caption{\textit{Areas and approximate 1$\sigma$ uncertainties for the 4.20 $\micron$, 4.67 $\micron$, 4.90 $\micron$, and 5.17 $\micron$ spectral features, measured by JWST/NIRSpec, and the CO$_2$ triplet band, measured by IRTF/SpeX, for the trailing hemispheres of Ariel (Ar), Umbriel (Um), Titania (Ti), and Oberon (Ob), normalized to Ariel. These normalized measurements use the same colors as the corresponding measurements shown in Figure \ref{band_areas}. A global, non-unique model (offset exponential) has been fit to the data to visually illustrate the comparable trends for each of these features as a function of distance from Uranus. The overall trend in band and peak areas demonstrates that CO$_2$ and CO are concentrated on the trailing hemispheres of the inner moons, possibly more consistent with a charged particle irradiation origin and less so  localized geologic exposure of CO$_2$ on each moon.}}
            \label{band_areas_summary}
		\end{figure}

        The measurements (Tables \ref{measurements_pt1}, \ref{measurements_pt2}, and \ref{measurements_pt3}) and analyses (Figures \ref{band_areas} and \ref{band_areas_summary}) reported here demonstrate that most of the spectral features associated with CO$_2$ and CO ($\sim$2.9 -- 5.3 $\micron$) are stronger on the trailing hemispheres of all four moons and decrease in strength with increasing distance from Uranus (Figure \ref{band_areas_summary}). These results are consistent with the hemispherical and radial trends measured in ground-based reflectance data for the CO$_2$ triplet band (shown here in Figure \ref{band_areas}e; \citealt{grundy2003discovery, grundy2006distributions, cartwright2015UmoonCO2}). 

       	Unlike prior ground-based observations, however, the G395M data for the leading hemispheres of these moons also exhibit CO$_2$ features, in particular a strong absorption band centered near 4.27 $\micron$, resulting from the $\nu$$_3$ TO mode of $^{12}$CO$_2$ ice. Indeed, the band area measurements reported here indicate that the 4.27 $\micron$ feature is stronger on the leading hemispheres of these moons compared to their trailing sides, unlike all other measurements of carbon oxides. Curiously, when comparing the four moons, Umbriel's leading hemisphere exhibits the strongest 4.27 $\micron$ band, unlike all other CO$_2$-related features that are significantly stronger on Ariel's trailing side, in both ground-based and JWST/NIRSpec datasets (Figure \ref{band_areas}). Additionally, all other measured CO$_2$ and CO features on Umbriel's leading hemisphere are weaker than those on its trailing side and weaker than those on the leading and trailing hemisphere of Ariel. 

        A plausible explanation for these apparent oddities is that the 4.20 $\micron$ scattering peaks observed on the trailing hemispheres of these moons, and Ariel's leading hemisphere, partially `fill-in' the adjacent 4.27 $\micron$ band, spuriously making these features appear weaker.  Possible evidence for such a fill-in effect can be seen in the trailing/leading spectral ratios, which exhibit large peaks at both 4.2 and 4.27 $\micron$ for all four moons (Figure \ref{spectral_ratios}). Photon penetration depths are extremely small in the wavelength range of the $\nu$$_3$ mode of $^{12}$CO$_2$ (see Figure 4 in \citealt{cartwright2024ArielJWST}), and light effectively interacts with only the outer surfaces of CO$_2$ ice grains, highlighting the challenges with using the 4.27 $\micron$ feature to diagnose the spectral properties of CO$_2$ ice. Consequently, we caution the reader from drawing strong conclusions from the 4.27 $\micron$ band strength trends reported here. We instead suggest that the spectral properties of weaker combination and overtone modes, outside the wavelength range of the strong $\nu$$_3$ mode for $^{12}$CO$_2$ ($\sim$4.1 -- 4.3 $\micron$), may provide more utility for understanding the spatial distribution and vertical stratification of CO$_2$ ice on the large Uranian moons, and perhaps other astrophysical environments that are enriched with crystalline CO$_2$ ice layers as well.

        As seen by JWST, the Uranian moons are point sources and all of the collected G395M data are disk-integrated. Because of the obliquity of the Uranus system (subsolar latitude $\sim$64$\degree$ at the time of these GO 1786 observations), the moons' observable disks included a fraction of their trailing hemispheres during the leading hemisphere observations and vice versa. Consequently, it is feasible that CO$_2$ located in deposits on their trailing hemispheres may contaminate the leading hemisphere observations reported here. Such `cross-hemispherical' contamination could hypothetically explain why CO$_2$ has yet to be confirmed on the leading hemispheres of these moons in ground-based data \citep{grundy2003discovery, grundy2006distributions, cartwright2015UmoonCO2}, which were mostly collected in late southern summer and early northern spring at Uranus (30$\degree$S -- 30$\degree$N), when cross-hemispherical contamination between their leading and trailing sides was negligible. To investigate this possibility, we calculated the fraction of cross-hemispherical contributions for each GO 1786 observation (using a disk-area calculation provided in \citealt{holler2016TritonIRTF}), finding only $\sim$26.7$\%$ overlap in these moons' disk areas, with most of this overlap ($\sim$18.7$\%$) limited to subsolar latitudes $>$45$\degree$N, where CO$_2$ is predicted to be strongly depleted if not entirely absent \citep{grundy2006distributions, sori2017wunda, steckloff2022exosphere, menten2024Ariel}. 
        
        It therefore seems feasible that the CO$_2$ detected in GO 1786 observations collected over the leading hemispheres of these moons results from deposits physically located on their leading hemispheres and spatially distinct from CO$_2$ on their trailing sides. Consequently, the results reported here represent strong evidence for the presence of carbon oxides on the leading hemispheres of Umbriel, Titania, and Oberon, further highlighting the ubiquity of CO$_2$ across the Uranus system. It is possible, however, that CO$_2$ ice deposits located on the leading hemispheres of these moons were initially exposed, or radiolytically formed, on their trailing sides, before migrating in response to ongoing seasonal sublimation-condensation cycles. Indeed, a global layer of CO$_2$ at least a few mm thick may migrate across the surfaces of these moons each Uranian year \citep{steckloff2022exosphere}, which would provide more than enough CO$_2$ to explain the detected spectral signature of this molecule on their leading hemispheres.  

    \subsection{CO$_2$ frost layer(s) mantling the regoliths of the Uranian moons?}

        Prior work indicates that exposed CO$_2$ layers are at least a few mm thick on Ariel's trailing hemisphere, possibly $\gtrsim$10 mm thick (see Figure 4 in \citealt{cartwright2024ArielJWST}). Based on existing laboratory data of thick CO$_2$ ice deposits, we expected to find near-saturated $\nu$$_3$ modes between 4.2 and 4.3 $\micron$ \citep{quirico1997near, quirico1997CO2SSHADE, hansen1997spectral}. For example, Neptune's moon Triton, which shows a wealth of weak CO$_2$ ice combination and overtone modes \citep[e.g.,][]{cruikshank1993TritonCOCO2, grundy2010TritonIRTF, holler2016TritonIRTF} similar to Ariel, exhibits near total absorption between $\sim$4.2 and 4.3 $\micron$ \citep{wong2025JWSTTriton}, consistent with the strong $\nu$$_3$ mode of CO$_2$, with likely contributions from an N$\equiv$N stretching mode exhibited by N$_2$ ice, centered near 4.25 $\micron$ \citep[e.g.,][]{dilella1986N2forbidden}. In contrast, this wavelength range is characterized by prominent scattering peaks ($\sim$4.16 -- 4.24 $\micron$) and relatively weak absorption bands ($\sim$4.24 and 4.30 $\micron$) on the trailing hemispheres of the large Uranian moons, as discussed in section 4.1. Based on the laboratory data presented here, CO$_2$ ice veneers can express large scattering peaks when condensed onto rough surfaces, as seen in our spectra of  CO$_2$ condensed onto organic residues and onto Infragold substrates (Figure \ref{CO2_4.0-4.6}). In contrast, CO$_2$ ice deposits on Triton may instead be dominated by geologically-thick outcrops of exposed `bed ice,' as suggested by a prior ground-based study \citep{holler2016TritonIRTF}. A possibly more relevant spectral analog for the Uranian moons may be found at Mars, where transient CO$_2$ frost clouds express prominent scattering peaks between 4.2 and 4.3 $\micron$ \citep[e.g.,][]{herr1970MarsCO2clouds, montmessin2007MarsCO2clouds, mangan2017MarsCO2clouds, aoki2018MarsCO2clouds}, as well as a host of other CO$_2$ features, including a prominent 3.33 $\micron$ band \citep{bell1996CO2clouds}. 
        
        Inspired by CO$_2$ frost clouds at Mars, we offer the following `working model' that may be able to explain the observed spectral properties of CO$_2$ on the Uranian moons. Perhaps the surfaces of the Uranian moons are coated by a highly porous veneer of CO$_2$ frost (perhaps 1 -- 10 $\micron$ thick), capping a more compacted layer of CO$_2$ ice, perhaps mixed with CO$_2$ clathrates, (few to 10's mm thick). These CO$_2$ layers mantle H$_2$O ice and dark regolith materials, possibly consistent with the low to moderate albedos and gray-toned surfaces imaged by V2 at each of these moons \citep[e.g.,][]{smith1986voyager, helfenstein1989evidence, helfenstein1991oberon, buratti1991comparative}. Such a layering of CO$_2$ is likely transient, especially the exposed CO$_2$ frost veneer, migrating in response to subsolar heating, as predicted by volatile transport models \citep{grundy2006distributions, sori2017wunda, steckloff2022exosphere, menten2024Ariel}. The seasonal migration-condensation cycling of CO$_2$ would continually refresh its spectral properties and likely help preserve the hypothesized high porosity veneer, which in turn could enhance scattering and contribute to the prominent 4.2 $\micron$ scattering peaks observed by JWST/NIRSpec. 
        
        Supporting this scenario, ground-based visible wavelength polarimetry data suggest that the large Uranian moons are mantled by a highly porous ($\sim$95$\%$), crumbly layer of fine-grained material ($\sim$1 $\micron$ diameters) of unknown composition \citep{afanasiev2014polarimetry}. Radiative transfer modeling of ground-based NIR observations also support the presence of small grains mantling the Uranian moons \citep{cartwright2020UmoonIRAC}, and suggest that at least two CO$_2$ ice grain sizes (1 -- 10 $\micron$ and $\sim$50 $\micron$) are required to match the morphologies of the CO$_2$ triplet bands \citep{cartwright2015UmoonCO2, cartwright2022ArielCO2}. Furthermore, Titania's H$_2$O ice features are stronger than Ariel's H$_2$O ice features in the G395M data (Table \ref{measurements_pt3}), but in ground-based datasets, Ariel exhibits stronger H$_2$O ice features than the other large Uranian moons. This disparity likely results from differences in photon penetration depths into H$_2$O ice, with light penetrating to greater depths at NIR wavelengths probed from the ground ($\sim$0.8 -- 2.5 $\micron$; $\sim$0.05 -- 10 mm depths) compared to the wavelength range of the G395M ($\sim$2.9 -- 5.3 $\micron$; $\sim$0.001 -- 0.05 mm). Consequently, shorter wavelength spectra could be better sampling Ariel's regolith underlying an exposed CO$_2$ ice veneer, thereby contributing to Ariel's stronger H$_2$O ice spectral features in ground-based datasets compared to the G395M data, where relatively CO$_2$-poor Titania expresses stronger H$_2$O ice features.  
                
        Therefore, the combination of scattering peaks and weak overtone and combination modes for CO$_2$ ice, and the disparities in the spectral signature of H$_2$O ice in different wavelength ranges, are potentially consistent with our working model involving two stratified layers of CO$_2$ ice. This model assumes that there is an exposed veneer of small grains that promotes scattering, with a layer of larger CO$_2$ ice grains, and perhaps CO$_2$ clathrates, beneath that dominates contributions to weaker combination and overtone modes. Both CO$_2$ layers  may be partially transient, migrating in response to subsolar heating, especially from the constantly sunlit summer poles, promoting the growth of long-term cold traps at low latitudes, where diurnal heating variations dampen sublimation rates.

    \subsection{Sources of CO$_2$ on the Uranian moons}

        \textit{Radiolytic CO$_2$?} Since the discovery of the CO$_2$ triplet band on Ariel, Umbriel, Titania, and Oberon, the preferred hypothesis to explain the distribution of this molecule has been radiolytic production driven by charged particle bombardment of preexisting carbon-rich components mixed with H$_2$O ice in the regoliths of these moons, primarily on their trailing hemispheres  \citep{grundy2003discovery, grundy2006distributions, cartwright2015UmoonCO2}. In this scenario, CO$_2$ is generated radiolytically from carbon-bearing material mixed with H$_2$O ice in the moons' regoliths, primarily at high latitudes, and then migrates in response to subsolar heating to long-term cold traps, located primarily at low latitudes. Additionally, CO would also be radiolytically generated, primarily from continued radiation processing of the newly formed CO$_2$ molecules. Such a combination of continual radiolytic production, seasonal volatile migration, and long-term sequestration of CO$_2$ (and perhaps CO) in cold traps could therefore explain how equatorial deposits of CO$_2$ ice gradually grow and remain quasi-stable over long timescales \citep{sori2017wunda, cartwright2022ArielCO2, menten2024Ariel}.  
        
        At first glance, the non-detection of the CO$_2$ triplet band on the innermost large moon Miranda \citep{bauer2002near, gourgeot2014near, cartwright2018Umoonred, decolibus2022MirandaH2O, decolibus2023MirandaNH3}, which is deeply embedded in Uranus' magnetosphere, would appear to be difficult to reconcile with a primarily radiolytic origin for CO$_2$. However, the considerably lower gravity of Miranda makes volatile loss via escape to space more effective compared to the larger moons. Indeed, thermodynamical models predict that $\sim$50$\%$ of migrating CO$_2$ molecules escape Miranda's surface before completing even one suborbital ballistic hop \citep{sori2017wunda}. As a result, the escape rate of CO$_2$ may outpace the radiolytic production rate of this molecule on Miranda, thereby preventing the buildup of thick deposits that can express CO$_2$ triplet bands. 

        As part of a different study, JWST/NIRSpec revealed the $\nu$$_3$ mode of solid-state CO$_2$ on Miranda, as well as in Uranus' rings and ring moons and on its far-flung irregular satellites \citep{belyakov2024irregularsJWST, belyakov2025smallsatsJWST}. When viewed in context with existing data of the largest moons, solid-state CO$_2$ is apparently widespread throughout the Uranian system. Uranus' rings also show hints of a 3.9 $\micron$ absorption band \citep{belyakov2024irregularsJWST, belyakov2025smallsatsJWST}, which might result from H$_2$CO$_3$ or other CO$_3$-bearing compounds. If these C-bearing constituents are present and mixed with H$_2$O ice, they may represent additional radiolytic products and/or some of the raw materials from which radiolytic CO$_2$ is being generated via interactions with Uranus' magnetosphere. Indeed, the G395M data of Ariel exhibit subtle features near 3.9 $\micron$ and between 5 and 5.1 $\micron$ (Figure \ref{spectral_ratios}), possibly resulting from H$_2$CO$_3$ \citep[e.g.,][]{moore1991H2CO3} and the free radical C$_2$O \citep[e.g.,][]{bennett2010CO2CO3}, respectively, which if present would be short-lived (especially C$_2$O) and likely diagnostic of ongoing radiation processing of Ariel's surface. Such moon-magnetosphere interactions would sputter CO$_2$ and other carbon oxides off Ariel’s surface, possibly contributing to an uncharacterized source of plasma detected near the moon's location in Uranus' magnetosphere \citep{cohen2023Arielplasma}, although CO$_2$ escaping from Miranda may also contribute to this plasma. Alternatively, these subtle bands may result from weak and unattributed CO$_2$ features, expressed by thick, crystalline ice-dominated deposits (see Figure \ref{CO2_phonon_modes} for an example of such an unattributed feature). 
        
        \textit{Native CO$_2$?} The spectral properties of CO$_2$ on Uranus' largest moons are largely consistent with crystalline CO$_2$ ice, exhibiting numerous subtle features only observed in thick layers of pure CO$_2$ ice \citep[e.g.,][]{hansen1997spectral, quirico1997CO2SSHADE, quirico1997near}. These subtle CO$_2$ features are not present in Uranus' rings and ring moons or its irregular satellites, hinting that at least some of the CO$_2$ on the large moons has an alternative source. In this scenario, CO$_2$ ice deposits may build up from passive outgassing, especially at the frigid temperatures of their winter poles (20 -- 30 K; \citealt{sori2017wunda}) where even hyper-volatile CO should condense. Such outgassing could be dominated by diffuse emissions from the interiors of these moons and not dramatic geyser eruptions. CO$_2$ clathrates may also be present and contribute to the multi-lobe structure of the $\nu$$_3$ mode of $^{13}$CO$_2$ (Figure \ref{13CO2_4.34-4.44µm}). Clathrate hydrates often form in high-pressure environments, such as on Earth's seafloors, and if present on the Uranian moons, they could point to an internal origin for CO$_2$ \citep{cartwright2024ArielJWST}. Clathrates can, however, form at low temperatures (10 K) and pressures ($\sim$10$^{-10}$ mbar) consistent with conditions in the interstellar medium (ISM) \citep{ghosh2023CO2clathratesISM}, and their presence does not necessarily assure an internal origin for CO$_2$ at the Uranian moons. Although, even in this `ISM scenario,' the presence of CO$_2$ clathrates on the Uranian moons would suggest sufficient timescales to enable CO$_2$-H$_2$O interaction and co-condensation to form thermodynamically stable phases. Additionally, the results reported here are consistent with prior spectral analyses that suggested internally-derived carbonate minerals could be present on Uranus' moons \citep{cartwright2023Umbriel,cartwright2024ArielJWST}.      
    
        Although we speculate that some weak features and spectral structure in the G395M data collected over Ariel's trailing hemisphere may result from the common radiolytic products H$_2$CO$_3$ and C$_2$O, we found no evidence for another established radiolytic product, H$_2$O$_2$, on Ariel or the other moons. H$_2$O$_2$ has been detected on the icy Galilean moons Europa and Ganymede and Pluto's moon Charon, primarily attributed to moon-magnetosphere interactions at Jupiter \citep[e.g.,][]{carlson1999EuropaH2O2, trumbo2023GanymedeH2O2} and solar ultraviolet and interplanetary medium Lyman-$\alpha$ photons, solar wind, and galactic cosmic rays at Charon \citep{protopapa2024CharonJWST}. Another study, reporting optical wavelength spectra of Ariel, Umbriel, Titania, and Oberon, found no evidence for spectral features consistent with trapped O$_2$ near 577 and 625 nm \citep{decolibus2026Umoonsoptical}, which is a telltale constituent for ongoing radiolytic processing of the icy Galilean moons' surfaces \citep[e.g.,][]{calvin1996GmoonsO2, johnson1997GanymedeO2, spencer2002GmoonsO2, oza2025GmoonsO2}. However, many of these H$_2$O-CO$_2$ irradiation products have not been detected on the surfaces of icy moons embedded in Saturn's magnetosphere, indicating that spectral signatures of radiolysis may be more subtle on icy bodies outside of the intense Jovian magnetosphere. Unlike Saturn's mid-sized, inner moons \citep{hendrix2018SmoonsHST}, the inner Uranian moons Ariel and Umbriel show no evidence for preferential radiation-darkening of their trailing hemispheres at ultraviolet and visible wavelengths \citep{soto2025UmoonsHST}, further supporting a scenario where radiation processing is not a dominant chemical process at these moons.
        
        \textit{Both native and radiolytic CO$_2$?} On balance, it seems plausible that solid-state CO$_2$ in the Uranus system has multiple origins, including radiolytic production from C-bearing material mixed with H$_2$O ice and geologic exposure and emplacement of native CO$_2$ ice and clathrates, perhaps mixed with other compounds that are hypothesized to be present and internally-derived on the large moons, such as carbonates \citep{cartwright2023Umbriel, cartwright2024ArielJWST} and NH$_3$ and NH$_4$-bearing species \citep{bauer2002near, cartwright2018Umoonred, cartwright2020ArielNH3, cartwright2023Umbriel, decolibus2023MirandaNH3}. Possibly analogous to this dual origin scenario for CO$_2$ on Uranus' moons, the spectral properties of CO$_2$ observed on the warmer surfaces of Jupiter's Galilean satellites have also been interpreted to support both radiolytic and endogenic origins. Data returned by the Near Infrared Mapping Spectrometer (NIMS) on the Galileo spacecraft measured $^{12}$CO$_2$ $\nu$$_3$ features with central wavelengths shifted to 4.25-4.26 $\micron$ on Europa, Ganymede, and Callisto, potentially consistent with radiolytically-produced CO$_2$ that is trapped or `complexed' within hydrated salts and/or organic residues mixed with H$_2$O-bearing minerals in which it is generated \citep[e.g.,][]{mccord1997NIMS, mccord1998NIMS, cartwright2024CallistoJWST}. Conversely, the observed spatial association between CO$_2$ and geologic features like craters on Callisto has been attributed to native CO$_2$, exposed by impact events \citep{hibbitts2000CallistoNIMS, hibbitts2002CallistoCO2}. Similarly, analyses of JWST/NIRSpec data of Europa suggest that some of the CO$_2$ on its surface is likely derived from its interior and exposed in chaos-dominated regions like Tara and Powys regiones \citep{trumbo2023EuropaCO2, villanueva2023EuropaJWST, cartwright2025EuropaJWST}. 
        
        Reflectance spectra collected over Europa's Tara and Powys regiones also exhibit 3.505 $\micron$ features, attributed to H$_2$O$_2$ formed via irradiation of H$_2$O ice \citep[e.g.,][]{carlson1999EuropaH2O2}. Laboratory experiments have demonstrated that the presence of a minor amount of CO$_2$ mixed with H$_2$O can enhance the formation of H$_2$O$_2$ \citep{mamo2025H2O2}, highlighting the chemical interplay between endogenically-derived CO$_2$, H$_2$O, and charged particles within the Jovian magnetosphere at Europa. Although rich in H$_2$O and CO$_2$, we find no compelling evidence for H$_2$O$_2$ on the Uranian moons, consistent with prior work for Ariel \citep{cartwright2024ArielJWST}, suggesting that surface radiation processing may be substantially weaker at Uranus. Alternatively, perhaps a cap of seasonally-mobile CO$_2$ effectively absorbs trapped charged particles, thereby shielding underlying H$_2$O ice from interactions with Uranus' magnetosphere and stymieing H$_2$O$_2$ production.  If such a cap exists, charged particle sputtering of CO$_2$ from this layer could be significant \citep{raut2013CO3}, hypothetically contributing to predicted, but yet-to-be-detected, exospheres at Ariel and the other moons \citep[e.g.,][]{steckloff2022exosphere,raut2025Arielexosphere}. Furthermore, the possible presence of carbonates supports endogenically-derived carbon oxides on these moons, which could feasibly include outgassed CO$_2$, as hypothesized for Europa \citep{trumbo2023EuropaCO2}, and would imply formation in the presence of liquid water \citep[e.g.,][]{carrozzo2018nature}. Carbonates could have formed in the interiors of the Uranian moons from interactions between melted H$_2$O ice and dissolved CO$_2$, mixed with silicate minerals \citep{glein2020EnceladusCO3}. If CO$_2$ was accreted by these moons, its presence would mean that temperatures in the circum-uranian disk (or subnebula) were below the condensation temperature of CO$_2$ ice when they formed (a representative upper limit is $\sim$80 K; \citealt{mousis2009clathration}). Of note, dust-delivered carbonate grains might also contribute to Callisto's 4 $\micron$ band \citep{cartwright2024CallistoJWST,sharkey2025JWSTirregulars}, perhaps complicating the endogenic origin story for CO$_3$-bearing species at Uranus.  

        Unraveling whether the primary source of CO$_2$ is \textit{in situ} production via radiation processing of regolith materials or passive outgassing and geologic exposure on the large moons is challenging using JWST and similar standoff platforms. Measurements made by instruments on an orbiter would vastly expand our knowledge of these carbon oxide-rich moons and the origin and fate of CO$_2$ across the broader Uranus system. In particular, a near infrared mapping spectrometer, dust analyzer, and ion and neutral mass spectrometer would be an ideal combination of instruments for constraining possible subsurface-surface-exosphere exchange of carbon oxides at Ariel and the other Uranian moons \citep{cartwright2021sciencecase}. A similar combination of instruments will be leveraged by the Europa Clipper spacecraft to understand material cycling on Jupiter's moon Europa \citep{becker2024Clippercomp}. 

	\section{Summary and Conclusions} 
				
	    We analyzed JWST/NIRSpec reflectance spectra collected over the leading and trailing hemispheres of Ariel, Umbriel, Titania, and Oberon, revealing prominent CO$_2$ ice and CO ice absorption features and scattering peaks, in particular on the inner moons, Ariel and Umbriel (Figures \ref{G395M_spectra}, \ref{spectra_zoomin}). We detected a suite of other spectral features that have previously been suggested to result from CO$_2$ clathrate hydrates, carbonates, carbon chain oxides, and nitriles \citep{cartwright2024ArielJWST}. Through comparison to laboratory data of CO$_2$ ice samples measured under a range of conditions relevant to the Uranus system, we determined that many of the spectral features between 4.8 and 5.25 $\micron$ can be attributed to collective oscillations across the crystalline lattice of thick CO$_2$ ice deposits (Figure \ref{CO2_phonon_modes}). Furthermore, the multi-lobe scattering peaks expressed by the $\nu$$_3$ mode of $^{12}$CO$_2$ on Ariel can be crudely approximated by a layer of CO$_2$ ice condensed on irradiated organic residues, as well as CO$_2$ deposited directly onto an Infragold substrate (Figure \ref{CO2_4.0-4.6}), highlighting the likely importance of surface roughness and grain boundaries for enhancing scattering. These same laboratory data can approximate the multi-lobe structure expressed by the $\nu$$_3$ mode for $^{13}$CO$_2$ ice on Ariel and Umbriel (Figure \ref{13CO2_4.34-4.52µm}), in particular when linearly mixed with a CO$_2$ clathrate spectrum (Figure \ref{13CO2_4.34-4.44µm}) presented in a prior study \citep{oancea2012CO2clathrates}. Laboratory spectra of CO$_2$ ice condensed on a meteorite simulant (amorphous MgFeSiO$_4$; \citealt{suhasaria2025CO2dust}) provides a less ideal match to the G395M data, but may  contribute to Umbriel's broad $^{13}$CO$_2$ feature (Figure \ref{13CO2_Suhasaria}). 
        
        Notable exceptions to this story include the 4.02 $\micron$ band (Figure \ref{cont-div_4um}) that is difficult to explain with existing laboratory datasets of CO$_2$ ice, and instead, is better matched by carbonates, similar to Jupiter's moon Callisto \citep{johnson2004radiation, cartwright2024CallistoJWST}. Our results also support the hypothesis that a 4.59 $\micron$ feature on Ariel and Umbriel could result from OCN$^-$, albeit a weak and poorly characterized shoulder feature seen in some laboratory data of CO$_2$ ice \citep{hansen1997spectral} may contribute to this band as well. To more fully understand the spectral properties of CO$_2$, CO, and other carbon oxides on Uranus' moons will require future laboratory experiments conducted under conditions relevant to the Uranian system, in particular for proper identification of species that express spectral features between 4.4 and 4.8 $\micron$ (Figure \ref{spectra_zoomin}). 
        
        We explored whether CO$_2$ is dominated by radiolytic production mechanisms or internally-derived components on Uranus' large moons, finding that the most likely outcome is a combination of both externally-derived and internally-sourced carbon oxides. The distribution of CO$_2$ and CO, concentrated on the trailing sides, primarily of the inner moons Ariel and Umbriel, would appear to be consistent with formation from charged particle bombardment, similar to the explanation for the distribution of CO$_2$ at the trailing hemisphere of Callisto \citep[e.g.,][]{hibbitts2000CallistoNIMS, cartwright2024CallistoJWST}. However, we found no trace of the common radiolytic product H$_2$O$_2$, whose production is enhanced when low levels of CO$_2$ are mixed with irradiated H$_2$O ice substrates \citep{mamo2025H2O2}, suggesting that radiation processing may be of secondary importance at the orbits of Uranus' large moons. The potential presence of CO$_2$ clathrates and carbonates, perhaps supplied by geologic processes, also hints at the presence of internally-derived carbon oxides on these moons. Furthermore, JWST has now detected CO$_2$ across the Uranian system, from its rings and ring moons to the irregular satellites \citep{belyakov2024irregularsJWST, belyakov2025smallsatsJWST}, demonstrating the ubiquitous presence of this molecule and further confounding a simple `one size fits all' origin story. 
        
        Based on the steadily growing stockpile of spectral evidence collected by JWST, and other platforms over the past several decades, we postulate that a dual origin scenario is required to explain the presence of CO$_2$ on the large moons of Uranus. By analogy, a variety of studies have found evidence for both geologically exposed and radiolytically produced CO$_2$ on the Galilean satellites \citep[e.g.,][]{mccord1998NIMS, hibbitts2002CallistoCO2, hibbitts2003GanymedeCO2, cartwright2024CallistoJWST, cartwright2025EuropaJWST}, highlighting the different possible origins for this molecule on other icy moons. To fully explore the origin(s) of CO$_2$ and other carbon oxides will likely require data collected by a Uranus orbiter to assess the spatial relationships between CO$_2$-rich deposits and specific geologic features and terrains and to more precisely trace time-varying interactions between Uranus' magnetosphere and its rings and moons \citep{national2022PSADS}. 
		
	\section{Acknowledgments} 

         This work is based [in part] on observations made with the NASA/ESA/CSA James Webb Space Telescope. The data were obtained from the Mikulski Archive for Space Telescopes at the Space Telescope Science Institute, which is operated by the Association of Universities for Research in Astronomy, Inc., under NASA contract NAS 5-03127 for JWST. These observations are associated with GO program 1786. Support for GO program 1786 was provided by NASA through a grant from the Space Telescope Science Institute, which is operated by the Association of Universities for Research in Astronomy, Inc., under NASA contract NAS 5-03127.
         We thank O. Mivumbi, K. Guernon, C. Lantz and L. Bouvet for help and support with INGMAR. INGMAR is a joint IAS-IJCLab facility funded by the P2IO LabEx (ANR-10-LABX-0038) in the framework Investissements d’Avenir (ANR-11-IDEX-0003-01). R.B. and S.C. acknowledge support from CNES (France) as part of their contributions to the JWST mission. Eric Quirico acknowledges financial support from the Centre National d’Études Spatiales (France) for the MAJIS/JUICE project. C.R.G. was supported by SwRI grant 15-R6440.		
	
\bibliographystyle{aasjournal}
\bibliography{References_2026}

\end{document}